\documentclass[lettersize,journal]{IEEEtran}
\usepackage{amsmath,amsfonts}
\usepackage{algorithmic}
\usepackage{algorithm}
\usepackage{array}
\usepackage[caption=false,font=normalsize,labelfont=sf,textfont=sf]{subfig}
\usepackage{changes}
\usepackage{textcomp}
\usepackage{stfloats}
\usepackage{url}
\usepackage{verbatim}
\usepackage{graphicx}
\usepackage{xcolor}
\usepackage{siunitx}
\usepackage{cite}
\usepackage{multicol}         
\usepackage{multirow}         
\usepackage{adjustbox}
\usepackage{booktabs}
\usepackage{amssymb}
\usepackage{wrapfig}
\usepackage{nomencl}

\makenomenclature

\begin{document}

\title{Massively Digitized Power Grid: Opportunities and Challenges of Use-inspired AI}

\author{Le Xie,~\IEEEmembership{Fellow,~IEEE}, Xiangtian Zheng,~\IEEEmembership{Student Member,~IEEE}, Yannan Sun,~\IEEEmembership{Senior Member,~IEEE}, \\Tong Huang,~\IEEEmembership{Member,~IEEE}, and Tony Bruton,~\IEEEmembership{Member,~IEEE}
\thanks{L. Xie and X. Zheng are with the Department of Electrical and Computer Engineering at Texas A\&M University (email: le.xie@tamu.edu, zxt0515@tamu.edu). T.Huang is with the Laboratory for Information and Decision Systems at Massachusetts Institute of Technology (tongh@mit.edu). Y. Sun and T. Bruton are with Oncor Electric Delivery (email: Yannan.Sun@oncor.com, Tony.Bruton@oncor.com).\\
The work of L. Xie, X. Zheng, and T. Huang was supported in part by the U.S. Department of Energy’s Office of Energy Efficiency and Renewable Energy (EERE) through the Solar Energy Technologies Office (SETO) under Grant DE-EE0009031, and in part by the National Science Foundation under Grant OAC-1934675, ECCS-2035688, and ECCS-1611301.}}

\markboth{Proceedings of the IEEE}%
{Xie \MakeLowercase{\emph{et al.}}: Massively Digitized Power Grid}


\maketitle

\begin{abstract}
This article presents a use-inspired perspective of the opportunities and challenges in a massively digitized power grid. It argues that the intricate interplay of data availability, computing capability, and artificial intelligence (AI) algorithm development are the three key factors driving the adoption of digitized solutions in the power grid. The impact of these three factors on critical functions of power system operation and planning practices are reviewed and illustrated with industrial practice case studies. Open challenges and research opportunities for data, computing, and AI algorithms are articulated within the context of the power industry's tremendous decarbonization efforts.
\end{abstract}

\begin{IEEEkeywords}
Decarbonization, power grid, data-driven algorithms, machine learning, artificial intelligence, industry use cases
\end{IEEEkeywords}

\nomenclature{AC}{Alternating current}
\nomenclature{AI}{Artificial intelligence}
\nomenclature{DC}{Direct current}
\nomenclature{DER}{Distributed energy resource}
\nomenclature{EMS}{Energy management system}
\nomenclature{EV}{Electric vehicle}
\nomenclature{PV}{Photovoltaic}
\nomenclature{SCADA}{Supervisory control and data acquisition}
\nomenclature{AGC}{Automatic generation control}
\nomenclature{SE}{State estimation}
\nomenclature{RTU}{Remote terminal unit}
\nomenclature{DMS}{Distribution management system}
\nomenclature{SSA}{Static security analysis}
\nomenclature{DSA}{Dynamic security analysis}
\nomenclature{FTR}{Financial transmission right}
\nomenclature{UC}{Unit commitment}
\nomenclature{ED}{Economic dispatch}
\nomenclature{OPF}{Optimal power flow}
\nomenclature{LMP}{Local marginal price}
\nomenclature{IBR}{Inverter-based resource}
\nomenclature{IoT}{Internet of things}
\nomenclature{PMU}{Phasor measurement unit}
\nomenclature{DFR}{Digital fault recorder}
\nomenclature{SOE}{Sequence of events}
\nomenclature{AMI}{Advanced metering infrastructure}
\nomenclature{FDR}{Frequency disturbance recorder}
\nomenclature{CEII}{Critical energy/electric infrastructure information}
\nomenclature{HIL}{Hardware-in-loop}
\nomenclature{GPU}{Graphic processing unit}
\nomenclature{ASIC}{Application-specific integrated circuit}
\nomenclature{BLAS}{Basic linear algebra subroutine}
\nomenclature{MIO}{Mixed integer optimization}
\nomenclature{ARX}{Autoregressive with exogenous input}
\nomenclature{PCC}{Point of common coupling}
\nomenclature{NN}{Neural network}
\nomenclature{SMT}{Satisfiability modulo theory}
\nomenclature{RPCA}{Robust principle component analysis}
\nomenclature{RL}{Reinforcement learning}
\nomenclature{MDP}{Markov decision process}
\nomenclature{HVAC}{Heating, ventilation, and air conditioning}
\nomenclature{RIC}{Residential/Industrial/Commercial}
\nomenclature{ELM}{Extreme learning machine}
\nomenclature{LSTM}{Long short term memory}
\nomenclature{CNN}{Convolutional neural network}
\nomenclature{KNN}{K-nearest neighbors}
\nomenclature{PCA}{Principle component analysis}
\nomenclature{SVR}{Support vector regression}
\nomenclature{RF}{Random forest}
\nomenclature{SVM}{Support vector machine}
\printnomenclature

\section{Introduction}\label{sec:introduction}
Digitization of the electric power grid, which broadly refers to the deployment of sensing, communication, and computational capabilities, has been an integral part of the electrification process over the past century and is a key enabling factor that drives power grid transformation by spreading its outreach vertically over plants, transmission grids, distribution grids, and end-use customers. As data availability and computing capacity continue to grow, large-scale power grids are built and operated with very high levels of reliability and efficiency, providing electricity services to billions of customers. 
The state of today's power grids in the United States (U.S.) can be summarized in three aspects: (i) for system reliability, the average duration of annual electric power interruptions in the U.S. varied from 3 to 8 hours in the period between 2013 and 2020~\cite{EIA_average_outage_hours};
(ii) for cost of electricity, the average wholesale electricity price across the U.S. varied from \$30 to \$60 per MWh in the period between 2016 and 2021~\cite{wholesale_price_mean};
and (iii) for carbon footprint, electricity generation in the U.S. produced an average of about 0.4 kilograms of carbon dioxide emissions per kWh in 2020~\cite{carbon_average}.


In response to climate change, which has emerged as a global concern, rapid decarbonization is imperative to reduce carbon emission, a quarter of which are contributed by the electricity sector. It is foreseeable that numerous decarbonization measures will cause profound changes in the electricity sector in the next few decades~\cite{rogelj2018mitigation}. Such changes have two major drivers: (i) the energy portfolio transition from high-carbon to low/zero-carbon generation sources, such as hydrogen, nuclear, wind and solar-based commercial generation units and distributed energy resources (DERs), and (ii) electrification in other sectors, including construction, transportation and other infrastructure systems. Deepening penetration of intermittent resources, such as wind farms and solar photovoltaic (PV), is introducing more variability and uncertainty. The proliferation of power electronics-based inverters is changing system dynamic characteristics. Increasing numbers of DERs at grid edge are strengthening the interaction between transmission and distribution systems. Rapid expansion of electric vehicles (EVs) will lead to substantial changes in electricity demand patterns.
Therefore, it is imperative for the grid operators to adopt a more flexible and risk-aware approach. Given the massive data availability and computing capacity provided by digitized power grids, data-driven artificial intelligence (AI) methods are feasible solutions for complementing traditional model-based approaches to address these complex emerging challenges.


From a broader economic perspective, AI has transformed a variety of domains over the past decade~\cite{ai100_2021}, including language processing~\cite{qiu2020pre}, speech recognition~\cite{nassif2019speech}, facial recognition~\cite{facial_recognization}, real-time object detection~\cite{redmon2018yolov3}, multiplayer game~\cite{vinyals2019grandmaster,silver2017mastering,silver2018general}, recommendation system~\cite{zhang2021artificial}, intelligent robotics~\cite{boston_dynamics_spot,boston_dynamics_atlas,agility_robotics}, driving assistant system~\cite{tesla_autopilot}, disease diagnosis~\cite{rajpurkar2017chexnet}, drug discovery~\cite{paul2021artificial}, finance~\cite{wu2020towards}, and others. 
We attribute such unprecedented success of AI as an intricate interplay between three factors, namely, \textit{massive data acquisition}, \textit{high computing performance}, and \textit{advanced AI algorithms}~\cite{bertsunas_youtube,openai,duchesne2020recent}.
The availability of data from heterogeneous resources has been increasing at an unprecedented rate ~\cite{oussous2018big,faroukhi2020big,agrahari2017review} and provides fuel for developing AI-based\added{,} data-driven applications for valuable knowledge extraction in wide-range domains.
In addition, remarkable improvements in computing performance have enabled a variety of practical large-scale AI models, credited to the collective advances in hardware, software, and computing architecture~\cite{leiserson2020there}.
Alongside rapidly-growing AI infrastructure that provides massive data and computing capacity, numerous advanced AI algorithms have been developed in the past decade. State-of-the-art performance on benchmark datasets for tasks in multiple research fields has been improved by pre-trained models~\cite{yang2019xlnet,devlin2018bert,szegedy2017inception,radford2018improving} and novel AI model architectures~\cite{krizhevsky2012imagenet,simonyan2014very,he2016deep,vaswani2017attention}.

\begin{figure}[hptb!]
\centering
\includegraphics[width=\columnwidth]{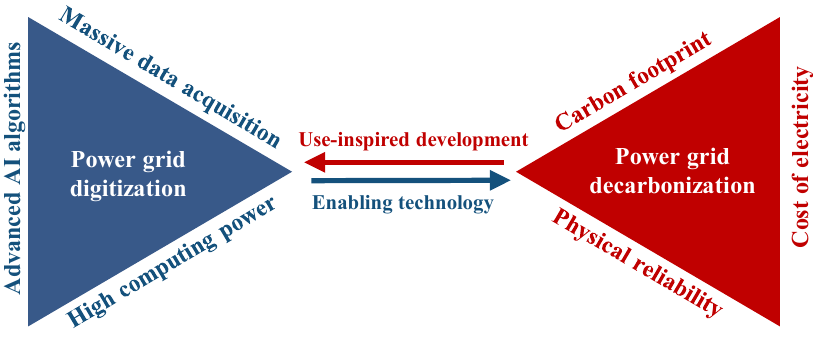}
\caption{Tri-factors of digitization are enabling technologies that facilitate the process towards power grid decarbonization while simultaneously meeting requirements in the aspects of reliability, cost of electricity, and carbon emission, while power grid decarbonization steers use-inspired development of power grid digitization.}
\label{fig:trifactor}
\end{figure}

Given the widespread success of AI applications, the development and deployment of interpretable, robust, and scalable AI may help to accommodate the emerging changes brought by decarbonization, aiming to reduce carbon emission and meanwhile ``keep the lights on" in a reliable and economic way (Fig.~\ref{fig:trifactor}).
However, to facilitate the process towards decarbonization, many open questions persist in implementing practical AI approaches in digitized power grids, including domain-agnostic computing and AI advances, use-inspired AI algorithm development, and cyber-physical security and privacy in a massively digitized power grid. To this end, this paper aims to provide a comprehensive review of the state-of-the-art practice of power grid digitization transformation, which focuses on three backbone factors: \textit{data}, \textit{computing}, and \textit{algorithms}. Specifically, this paper provides a review of the recent progress in data acquisition, computing capability, and AI algorithms that are applicable to power systems. Successful industry use cases are introduced to illustrate applications of AI algorithms on large real-world data sets.

The rest of the paper is organized as follows. Section~\ref{sec:review of operation} provides an overview of power grid operation and planning practices, as well as the challenges posed by decarbonization.
Sections~\ref{sec:data_in_PS},~\ref{sec:computing_in_PS} and~\ref{sec:algorithm} provide a comprehensive review of data, computing, and algorithmic advances in power systems. Section~\ref{sec:use cases} provides an industry perspective on AI adoption. Finally, Section~\ref{sec:outlook} concludes the paper with remarks on future directions for power grid modernization. 

\section{Physical and Market Operations of Power Systems}\label{sec:review of operation}
\begin{figure*}[tb]
\centering
\includegraphics[width=\textwidth]{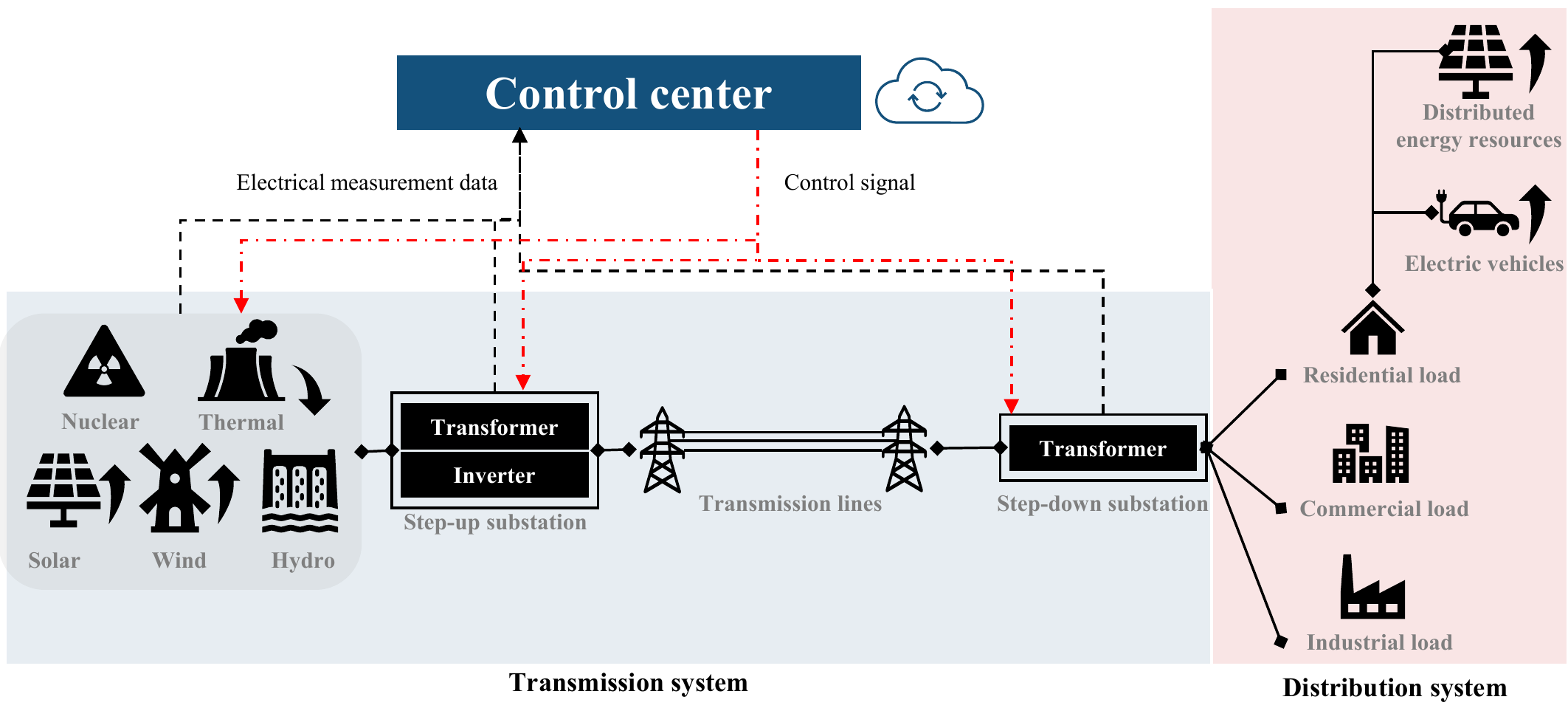}
\caption{Conceptual diagram of a modern power grid, consisting of transmission and distribution systems. The high-voltage transmission system consists mainly of generation, substation, and transmission lines. The low-voltage distribution system supplies electricity to residential, commercial, and industrial load.  Decarbonization has promoted utility-scale renewable generation, distributed energy resources and electric vehicles, while reducing investment in thermal generation. Digitization has contributed to the reform and upgrade of control centers through the development of cloud data storage and computing and the deployment of massive digitized sensors across the grid.}
\label{fig:power_grid_diagram}
\end{figure*}

Modern power grids are being driven by strong momentum of decarbonization~\cite{mcdonald2019digitized} with decentralization and transportation electrification. Fig.~\ref{fig:power_grid_diagram} shows the brief conceptual diagram of a modern power grid, which can be separated into transmission and distribution systems. Transmission systems refer to bulk systems that have voltages higher than 66 kV and consist of generation, substation and transmission lines, which are usually operated by state-wide or cross-state system operators.
Distribution systems refer to close-to-users systems that have voltages lower than 33 kV and connect to residential, commercial and industrial load, which are usually operated by local utility companies. Power grid decarbonization is changing the energy portfolio in terms of generation resources, such as increasing commercial-size solar PV and wind farms in transmission systems, and DERs such as rooftop solar PV in distribution systems. Power electronics-based inverters are thus being deployed to convert electricity by renewables from direct current (DC) to alternating current (AC). Transportation electrification introduces a rapidly expanding number of electric vehicles into distribution systems.

\begin{figure*}[tb]
\centering
\includegraphics[width=\textwidth]{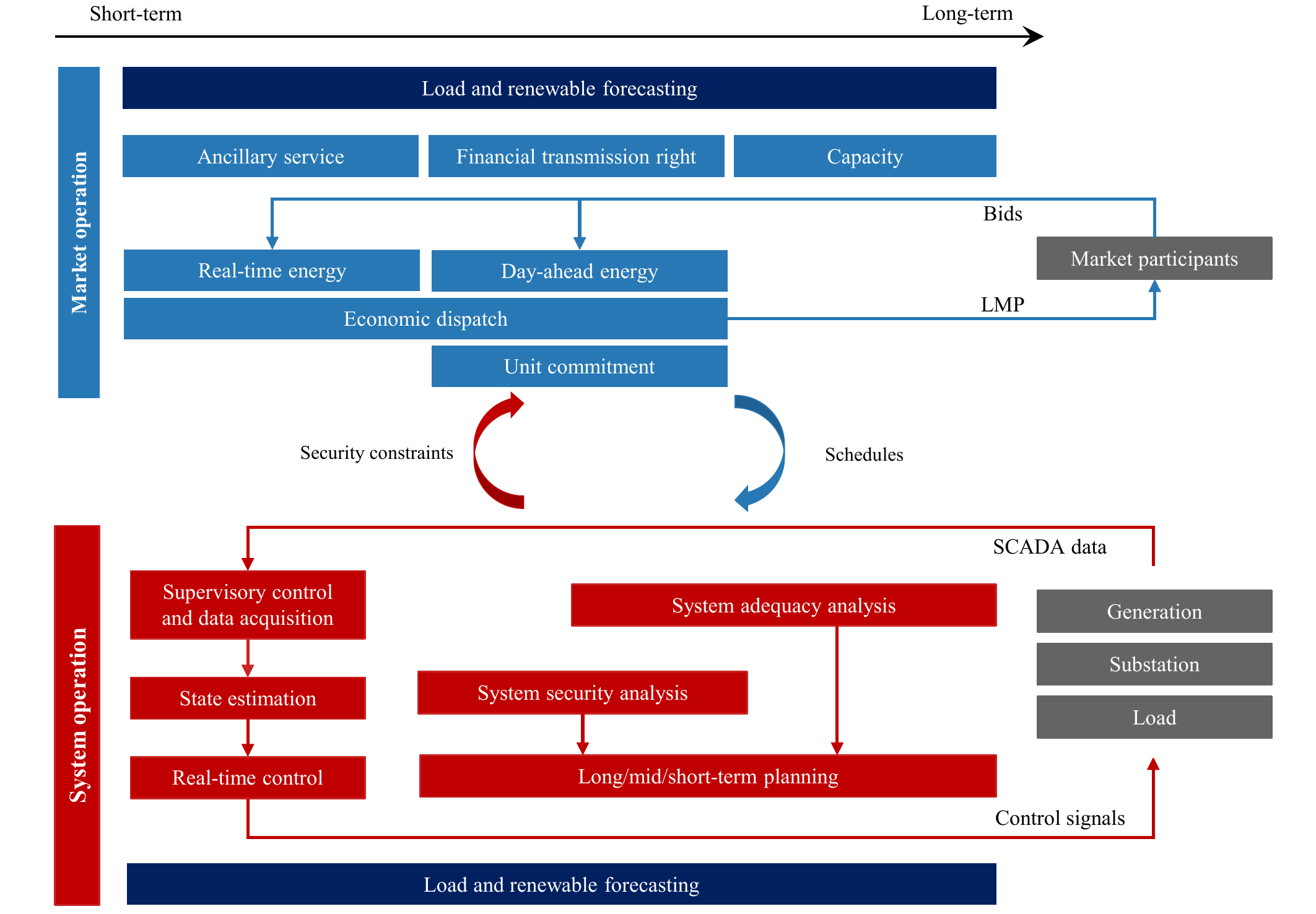}
\caption{Structure of physical and market operation in transmission systems.}
\label{fig:power_grid_EMMS}
\end{figure*}

The modern power system operations in high-voltage transmission systems can be broken down into two categories~\cite{wu2005power}. The first category is physical operations, which are responsible for the grid's physical security\footnote{Physical security in power systems refers to the ability to resist contingency disturbances, such as a transmission line short circuit and loss of system components.} and resource adequacy;\footnote{Resource adequacy in power systems refers to the ability to supply electricity that accommodates load variation, renewable uncertainty, and system component outages.} the second concerns market operation. Both physical and market operations are summarized in Fig.~\ref{fig:power_grid_EMMS}.

\subsection{Functions of Physical Operation and Planning}
Power system operation and planning fulfills the reliability of power systems via multiple functions including real-time monitoring, control, protection, and system reliability analysis.
A system-wide monitoring system collects and processes measurements, and presents intuitive information to system operators via visualization and alarming. A control system performs control actions either manually or by automated procedures. A protection system executes prescribed corrective measures upon detection of anomalies within targeted system components, which is achieved mainly by local sensors and actuators.  Reliability analysis provides instructions on decision making of multiple time horizons to guarantee the system within adequacy and security criteria. 

Load and renewable forecasting provides input for both system and market operation, by estimating uncertain net load and renewable generation of various projection horizons.
Load forecasting covers various prediction horizons spanning hours, days, weeks, months, and years ahead, whereas renewable forecasting provides only hours and days-ahead predictions. In real-world power grids, short-term load forecasting typically has high accuracy and renewable forecasting also has acceptable errors that can be mitigated by real-time operation of dispatchable resources.

Real-time monitoring and control are implemented mostly by \textit{energy management systems} (EMS) in the control center, the primary functional modules of which mainly include \textit{supervisory control and data acquisition} (SCADA), \textit{state estimation} (SE), and \textit{automatic generation control} (AGC). The SCADA system fulfills measurement acquisition and control telemetry through communication channels between the control center and \textit{remote terminal units} (RTUs), at the respective electrical station or device. Typically, the data acquisition function collects measurements every 2 to 10 seconds, of which the data stream is a key enabling factor for realizing other functionalities such as state estimation, real-time control, unit commitment, and economic dispatch. For accurate situational awareness of the system's current operation, function SE provides the steady-state estimation of system variables that are not directly observed in streaming SCADA data. As one of the major real-time control, primary and secondary generation control are implemented to (i) regulate load frequency, and (ii) balance power generation, load demand and cross-area interchange in real time. Droop-based generator governors that are responsible for primary control perform instantaneous power quality corrections before triggering protection relays. AGC, considered as secondary control, mitigates unavoidable errors of primary control by sending commands from the control center to participating generation units every 2 to 4 seconds~\cite{apostolopoulou2014automatic}.
Real-time protection is mainly implemented by protective relays that are equipped to critical assets, such as generation units and substations. In high-voltage transmission systems, protective relays should clear faults within several cycles\footnote{One cycle of a 60-Hz electric power system is about 16 ms.} to avoid further system deterioration.
Similarly, a distribution management system (DMS) enables real-time monitoring in the distribution system, with a few similar functions to EMS, such as SCADA and event analysis~\cite{cassel1993distribution}. It is worth noting that most field devices in the distribution systems are manually operated rather than remotely controlled, indicating a lower level of automation compared to the transmission system.

System reliability analysis entails adequacy, static security and dynamic security analysis~\cite{morison2004power,duchesne2020recent,allan2013reliability}. Security analysis focuses on the process of system state transitions initiated by reasonable disturbances such as short circuits and loss of system components. \textit{Static security analysis} (SSA) evaluates the viability of post-event equilibrium by calculating power flow or optimal power flow to check whether a power or voltage violation happens after an $N-1$ contingency.\footnote{The $N-1$ contingency refers to loss of a single system component, such as generation outage and transmission line tripping.} \textit{Dynamic security analysis} (DSA) evaluates the ability of the system to transition from one equilibrium to another post-event equilibrium within security criteria~\cite{duchesne2020recent} by simulating on system dynamic models. Adequacy analysis quantifies the system's capacity for sustainable supply that accommodates load variation, renewable uncertainty and system component outages by several manually defined indices. A typical method for adequacy and security analysis is numerical simulation. Due to time intensity, these reliability analysis methods tend to be impractical for real-time security control during contingencies. SSA and DSA are used in short-term scheduling, such as generation scheduling, which is performed daily or every few hours. Adequacy analysis and SSA are typically used for mid-term planning, such as facility maintenance, that is performed every several months to one year. Also, both adequacy and security analysis are used for long-term planning, which occurs annually or every few years.

\subsection{Functions of Market Operation}
Market operation in wholesale electricity markets aims to maximize social welfare while obeying physical constraints. Wholesale markets comporise day-ahead and real-time energy markets, capacity markets, financial transmission right (FTR) markets and ancillary service markets. Both day-ahead and real-time energy markets determine clearing prices based on bids from market participants, incorporating physical constraints and potential restrictions. Capacity markets ensure long-term system reliability. FTR markets entitle market participants to offset potential losses (hedge) related to the price risk of delivering energy to the grid. Ancillary service markets provide regulation and reserve.
\textit{Unit commitment} (UC) and \textit{economic dispatch} (ED) are two major security-constrained, bid-based mechanisms to handle the scheduling of generation and the management of system congestion. Both UC and ED are typically formulated as large-scale nonlinear/linear programming problems, known as \textit{optimal power flow} (OPF). Providing forecasted load and renewable as input, the UC function determines when and which generation units start up and shut down in day-ahead markets. The ED function calculates the power output of each committed generation unit and associated local marginal prices (LMPs). ED is performed to meet the day-ahead hourly forecasted load in day-ahead energy marketsas well as to meet the minute-ahead forecasted load every 5 to 10 minutes in real-time energy markets~\cite{stoft2002power}.

In today's distribution grids, the retail merket contains few centralized operation or scheduling functions, such as UC and ED, in the retail market. Given the proliferation of DERs in distribution grids such as distributed generation, interruptible load, and electricity storage, the retail market will involve system upgrades and reforms in the future to accommodate DER market participation, and to establish an appropriate mechanism of scheduling and compensation~\cite{haider2020toward}.

\subsection{Challenges of Decarbonizing Power Grid}

Renewable integration and transportation electrification at scale impose challenges on the paradigm of protection and control. The emergence of massive grid-following and grid-forming inverted-based resources (IBRs) may challenge the effectiveness and efficiency of the current central control frame due to the unknown impacts of electromagnetic dynamics and low inertia.
DERs at the grid edge may create bi-directional power flows that potentially incur malfunctions of the protective relays in distribution grids.
Besides, typical methods for adequacy and security analysis are numerical simulations that highly rely on grid models of multiple time scales, including electromagnetic dynamic (very fast), electromechanical dynamic (fast), and steady state (slow). However, system characteristics are being changed due to the proliferation of inverter-interfaced renewable resources and EVs in modern power grids, such as low inertia and deeper integration of transmission and distribution systems. These emerging system characteristics create a need for new requirements on the existing models to determine whether the system is within critical security criteria. For example, there is an urgent need for the study of several topics in order to handle the growing system complexity, including (i) electromagnetic transient models to reveal the fast dynamics by power electronic-based system components, (ii) system-level joint simulation between transmission and distribution models to reveal the increasing cross-system interaction, and (iii) cross-domain electricity-transportation models to incorporate the impacts of transportation networks on EVs.

Market operation also faces the challenge of managing potential market risks resulting from the variability and stochasticity of renewable generation~\cite{joskow2019challenges}. Strong uncertainty is a key obstacle to economic dispatch to (i) maintain system stability as tertiary frequency control and (ii) avoid unexpected renewable curtailment to the greatest possible extent to achieve decarbonization. Current wholesale markets may not be sufficiently prepared to accommodate increasingly frequent extreme weather events such as the 2021 Texas power outage event~\cite{ferc_ercot_2021} to prevent spiking price and mitigate energy scarcity.
Specifically, strong uncertainty regarding system net load and intermittent renewables generation in future grids will raise severe challenges for the accuracy and robustness of short-term load and renewable prediction. Deepening transportation electrification may also undermine the existing end-use and econometric models for medium and long-term load forecasting~\cite{feinberg2005load}.

The distribution system also faces a growing number of facility challenges. Aging power lines may limit maximum use of renewable energy sources, such as wind farms and utility scale solar, especially in less populated areas where large renewable energy installations are located. The utilization and availability of DERs installed in densely populated areas can be affected by frequent localized outages intermittently that may be recognized by the control center. Given stronger integration and correlation between transmission and distribution grids, facility outages, such as transformer failures, may cause wider impacts. Furthermore, in aiming to establish a competitive retail market in the distribution system, there multiple critical problems remain unsolved, such as LMP calculation and demand response modelling; however, these are beyond the scope of the paper.

Overall, the profound changes by decarbonization are posing and will continue to pose numerous challenges to all aspects of physical reliability and economics. Given massive data acquisition as the ``fuel" and high computing power as the ``engine," applying advanced data-driven AI-based approaches as an ``autopilot" have the potential to steer the vehicle forward in a flexible and risk-aware manner.
\section{Data Acquisition in Digitized Power Grids}\label{sec:data_in_PS}
In broad industry sectors, large-volume and heterogeneously structured data have been generated at an unprecedented rate by diverse resources since 2010~\cite{oussous2018big,faroukhi2020big,agrahari2017review}, such as Internet of Things (IoT) records, social media, smart devices, and healthcare systems. The availability of such tremendous volumes of data has facilitated numerous applications of valuable knowledge extraction in sectors~\cite{national2013frontiers} such as spanning manufacturing~\cite{hilbert2016big}, healthcare~\cite{dimitrov2016medical}, government~\cite{pencheva2020big}, retail~\cite{einav2014economics}, infrastructure~\cite{chen2016promises,zhou2016big,baccarelli2016energy}. In particular, numerous high-quality open-source training datasets~\cite{wikipedia_datasets} have been created to boost AI research in the aspects of model training, testing, calibration, and benchmarking.

Moving with the tide of digitizing power systems, the explosive growth of data resources has also created massive volumes of data in heterogeneous formats, including electrical measurements that span across grids vertically, such as sensors installed on grid-level components, smart meters  and smart appliances as well as non-electrical measurements, such as weather, social media, traffic and geographic information~\cite{Bhattarai2019Big}. These data have proven very valuable in many use cases such as asset assessment, operation planning, real-time monitoring, and protection~\cite{kezunovic2013role}. It is worth noting that these basic functionalities have distinct requirements for data quality in perspectives of data accuracy, latency, and sampling rate~\cite{huang2016data}.
This section will review data acquisition approaches of electrical measurements in the power grids.

\subsection{Real-world Measurements in Power Systems}
\subsubsection{Sensors in Transmission Systems}
SCADA systems, which have played an important role in transmission system operation, are capable of collecting facility information and sending control signals, which are implemented by the critical component(i.e. RTUs). SCADA systems collect asynchronous data on bus voltage magnitude as well as active and reactive power flows; the typical reporting rate is merely 1 sample per 2 to 6 seconds. The wide-range acquisition of SCADA data has facilitated remote monitoring and system operation automation. For example, the EMS at the control center is capable of estimating physical state variables that are not directly observable based on SCADA data alone. However, due to increasing system complexity and uncertainty, even this successful SCADA-based application is becoming inadequate.


Phasor measurement units (PMUs) have been deployed in the bulk transmission grid at an accelerated rate after the 2003 U.S. blackout~\cite{2003blackout}. PMUs are able to measure the voltage phasors\footnote{Phasors contain magnitude $A$ and phase angle $\phi$ of sinusoidal waveforms that can be expressed as $A\text{sin}(\omega t+\phi)$, where $\omega$ is $2\pi\times60$ rad/s in a 60-Hz system.} at the installed bus (typically substations) and current phasors of the lines connected, along with synchronized time stamps, for which the typical reporting rate is 30 or 60 samples per second. Compared to SCADA, PMUs' high accuracy of time stamps and sensing, low latency, and high sampling rate of PMU benefit basic functionalities to different degrees~\cite{hojabri2019comprehensive}: (i) more real-time control and protection applications become potentially implementable due to all of these advantages, such as remedial action schemes including grid islanding and short-term stability control; (ii) online system security analysis, such as disturbance detection and situational awareness, can be significantly improved due to low latency; (iii) system adequacy analysis for long-term planning, such as model calibration, can be improved due to high accuracy. However, it is worth noting that, due to several factors such as high costs and time consumption of installation, only around 2,500 production-grade PMUs have been installed across the North America transmission power grid~\cite{usman2019applications,NASPI_PMU_map_2017}.

Digital fault recorders (DFRs) capture and store transient data and sequence of events (SOE) data that can be used for various purposes such as protection scheme monitoring and fault diagnosis, which tend to be implemented offline. DFRs have three typical recording mechanisms: steady-state, low-speed and high-speed disturbance recording modes. The disturbance recording modes are usually triggered by signals from protection relays. The steady-state recording mode captures the min, max and mean values of phasors at a low sampling rate of 1 sample per 10 seconds to 1 hour. The low-speed disturbance mode aims to provide phasor-domain information of long-term and short-term disturbances at a sampling rate of 1 sample per 1 to 10 cycles. The high-speed disturbance mode aims to record instantaneous time-domain voltage and current measurements of transient faults at a sampling rate of hundreds of samples per cycle. 

\subsubsection{Sensors in Distribution Systems}
The rapid expansion of \emph{advanced metering infrastructure} (AMI) meters at grid edge has created massive amounts of residential electricity consumption data, typically at a rate of 1 sample every 1 or 5 minutes. For example, the Pacific Gas and Electric Company collects more than 3 terabytes of power data from 9 million smart meters across the grid in the territory, and the State Grid Corporation of China collects 200 terabytes of data per year~\cite{guo2018complex}.

SCADA in distribution systems has facilitated remote monitoring and automated operation in multiple aspects, such as substation, feeder, and end-user load control. In substation systems, SCADA gathers data including voltage magnitude, current magnitude and binary status of facilities such as switches, breakers, and transformers. In typical feeder systems, SCADA facilitates the collection of historical data from feeder status of devices such as controlled load break switch and reclosers. In end-user load, SCADA collects all meter data from the end users.

The frequency disturbance recorder (FDR), one of representative PMU applications in distribution systems, is a GPS-synchronized single-phase PMU at ordinary 120-volt wall outlets. FDRs have the advantages of low cost and high deployability; they can be deployed even at residential households and campuses~\cite{wang2007frequency}. Using hundreds of FDRs that have been strategically placed across the U.S., the frequency monitoring network FNET/GridEye~\cite{FNET} is able to provide visualized nation-wide frequency monitoring. 

\subsection{Artificially Generated Power System Data}
Artificially generated data are commonly used for power system research for two major reasons: (i) most real-world operational data are protected by policies such as Critical Energy/Electric Infrastructure Information (CEII) owing to confidentiality, and (ii) real-world measurement datasets of high-impact events are usually insufficient for data-driven model training, due to the reliability of real-world power grids, which ensures that high-impact events are rare. Alternatively, artificial data generation methods facilitate the gathering of arbitrary numbers of data samples under varying scenarios and conditions, including voltage, current, frequency, and even machine inner state measurements across grid models.
\subsubsection{Model-based Simulation}
\begin{figure}[tb]
\centering
\includegraphics[width=\columnwidth]{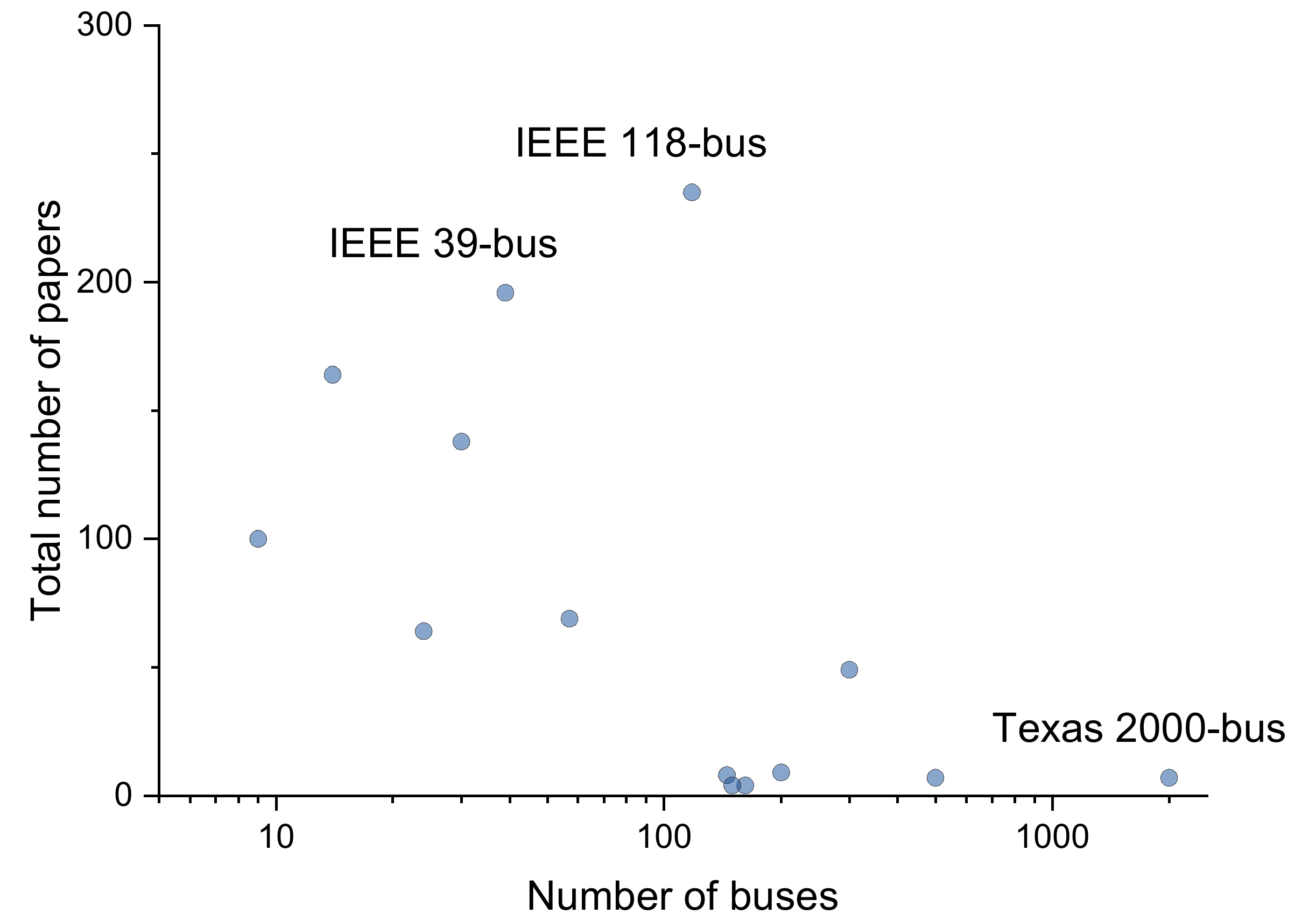}
\caption{Number of AI/ML/DL papers per simulation model of various system scales. Note that we only count typical open-source simulation models, including IEEE standard test cases and large-scale synthetic grids using Google Scholar advanced search among IEEE transaction papers from 2016 to 2021.}
\label{fig:stats_simulation_model}
\end{figure}

Model-based simulation is one of the most common data acquisition approaches for research and education purposes. Simulation models of transmission and distribution systems can be categorized into two major types: (i) small-scale standard systems and (ii) large-scale synthetic systems, which are available at~\cite{TAMU_test_systems}.
IEEE standard test systems are typically used for investigations such as algorithm assessment and power system analysis. 
Researchers have recently contributed to the creation of large-scale synthetic grid models \cite{birchfield2016grid} that possess realistic system characteristics. These large-scale synthetic grids have been used for analysis such as macro-scope energy portfolio transition~\cite{BTE1,BTE2} and quantitative assessment of measures against extreme events~\cite{wu2021open}.
For intuitive impression, we show the ``popularity" of simulation models in Fig.~\ref{fig:stats_simulation_model} by counting the number of corresponding IEEE transaction papers,\footnote{We use the Google Scholar advanced search to count IEEE transaction papers from 2016 to 2021, using several keywords such as ``IEEE Transactions on", ``machine learning", ``\#-bus system", ``power flow" and ``transient".} which are used for both machine learning model training and testing. It is clear that the most commonly used models for AI algorithm training, testing and calibration are the IEEE 39-bus and 118-bus systems, whereas the large-scale models are rarely adopted. Please refer to Tables~\ref{tab:load_modeling}-\ref{tab:grid_online} for other simulation models that are not included in Fig.~\ref{fig:stats_simulation_model}.
\subsubsection{Hardware Test Bed}
The development of hardware-in-loop (HIL) simulators has been used to support various types of research, including event detection, situational awareness, wide area monitoring and control, and cyber security~\cite{adhikari2012development}. HIL leverages the interface between a real-time software simulator and a hardware system to enable closed loop control~\cite{montoya2020advanced}. HIL may play an important role in electromagnetic transient simulation of electronics-rich power grids because of its ability to represent realistic very-fast dynamics.

This section gives an overview of data acquisition approaches in today's electric power grids. The rapid expansion of advanced sensors across systems and the development of simulation have facilitated massive data acquisition spanning multiple spatial and temporal scales and have further accelerated practical data-driven applications. Efforts to explore data-driven innovation, such as big data hubs~\cite{NSF_datahubs,NSF_south_datahubs}, have also promoted data-intensive research in the power system industry as well as academia and education. Despite these advances, there are several key challenges regarding the data for AI algorithms. 
First, in contrast to numerous datasets that have benefited broad AI communities, the lack of publicly accessible high-quality power datasets may be impeding the advancement of AI research in power systems. For example, insufficient data representativeness is one of the decisive factors for data-hungry AI methods. Real-world measurements cannot provide a sufficient volume of publicly available data due to confidentiality rules and strong grid reliability. Randomly sampled scenarios in simulation can generate massive amounts of data, but they do not necessarily guarantee representativeness;therefore they likely lead to unexpected training biases, which was demonstrated by the example of ACOPF scenario generation~\cite{joswig2021opf}.
Second, the feasibility of the proposed AI algorithms may be constrained by the current data acquisition system, as indicated by the data quality requirements of major power system applications~\cite{huang2016data}. For example, limited and inappropriate placement of high-sampling sensors that determine situational awareness for a specific task may confine advanced analysis and control including but not limited to practical applications of AI methods.
Third, although AI methods may offer unique creativity given cross-domain datasets, they require deep interdisciplinary knowledge and collaboration to identify useful combinations of heterogeneous datasets, which has been demonstrated by few AI-based canonical studies, such as automatic classification of distribution grid phases by camera imaging~\cite{sheinin2017computational} and comprehension COVID impacts on power sectors by mobile phone location data~\cite{ruan2020cross}.

\section{Computing in Digitized Power Grids}\label{sec:computing_in_PS}
Given sufficient available data resources, the implementation of data-driven applications in modern power grids faces computational burdens derived from large-volume, heterogeneous data. Such implementation is critical to handle the associated challenges, which include data streaming storage, querying and processing. This section will give an overview of state-of-the-art computing that has facilitated general AI, and will then introduce data streaming management systems and data processing platforms~\cite{guo2018complex,khan2021review} in power systems.

\subsection{Overview of State-of-the-art Computing for AI}
The remarkable improvement of computing performance is the key factor in the proliferation of AI, which is attributable to advances in hardware, software, and generic algorithms~\cite{leiserson2020there}. Quantum leaps in computing performance have yielded a variety of practical large-scale AI models, among which the amount of computation for model training has been increasing exponentially with a 3.4-month doubling period~\cite{openai,thompson2020computational}.
The rapid progress of hardware computing resources has been the main driver behind the development of AI models. Of particular note, the emergence of general purpose graphics processing units (GPUs)~\cite{NVIDIA_V100} and AI accelerator application-specific integrated circuits (ASICs), such as~\cite{google_TPU,merolla2014million,graphcore_ipu,nvidia_MCU}, are capable of dramatically accelerating AI model training.
In addition, AI-tailored software has been developed to exploit hardware computing resources~\cite{capra2020hardware}. For instance, basic linear algebra subroutine (BLAS) libraries, which were created decades ago~\cite{lawson1979algorithm,dongarra1990algorithm,golub2012matrix},  have been used to optimize common linear algebra operations that are recursively executed in deep neural networks~\cite{AMD_LibM,Intel_MKL,Nvidia_cuBLAS,opencl}. In particular, Nvidia GPUs, which are widely supported by mainstream deep learning framework~\cite{abadi2016tensorflow,jia2014caffe,paszke2017automatic}, have a highly optimized library cuDNN~\cite{Nvidia_cuDNN} enabling high-performance GPU acceleration.
The progress of generic algorithms has also improved computing performance, exhibiting enormous heterogeneity on problems of different types and sizes~\cite{sherry2021point}. It is worth noting that some large-size problems benefit just as much or even more from algorithmic improvement than from Moore's law. 
For instance, the total speedup of solving mixed integer optimizations (MIO) was 2.2 trillion times during the 25 years between 1991 and 2016~\cite{bertsunas_youtube}, of which a factor of 1.6 million is due to hardware speedup from 59.7 GFlop/s in 1993 to 93.0 PFlop/s in 2016; another factor of 1.4 million is due to software and algorithmic speedup from CPLEX 1.2 in 1991 to Gurobi 6.5 in 2015.

\subsection{Data Management Platforms in Power Grids}
Because power system security highly relies on real-time system operation and control, it is challenging to store and process real-time data streaming effectively and efficiently. Therefore, the building of real-time data streaming systems that mainly influence data latency is critical for the subsequent online data-driven applications including but not limited to AI-based methods. In contrast to traditional database management systems that use statistical data storage, data stream management systems usually store synopsis data (instead of the entire dataset) via processing in order to handle frequent queries and data update. We illustrate several of the most popular data stream management systems summarized in~\cite{guo2018complex}: Aurora~\cite{carney2002monitoring} has a good balance of accuracy, response time and resource utilization; TelegraphCQ~\cite{hellerstein2000adaptive} is mainly used for sensor networks, which involves a front end, a sharing storage, and a back end; STREAM~\cite{babcock2002models} has the advantage in situations of limited resources in that it can execute queries with high efficiency.


In particular, big data management platforms are being developed to accommodate multi-modal data storage and processing of unstructured heterogeneous data. Hadoop~\cite{shvachko2010hadoop} and Spark~\cite{zaharia2016apache} are two representative open-source designs for distributed data management. Hadoop is able to process massive heterogeneous data efficiently and economically by taking advantages of a programming model~\cite{dean2008mapreduce}, a distributed file system~\cite{ghemawat2003google}, and a distributed data storage system~\cite{chang2008bigtable}. Spark, on the other hand, leverages the technology of resilient distributed datasets\cite{zaharia2012resilient}, which is more suitable for recursive computational operations in machine learning-based applications. In terms of data management platforms that are suitable for power systems, several cases of solutions have been successful in facilitating energy efficiency. For example, CenterPoint Energy has handle streaming messages from intelligent grid devices and smart meters using an IBM-developed platform to improve system reliability~\cite{IBM_centerpoint}. For its part, Oncor Energy Delivery has developed AMI data-based predictive maintenance to reduce outages and guarantee sustainable supply enabled data platforms~\cite{IBM_Oncor}.



Because of power grid digitization, computing tasks in today's power grids have been shifted and evolved to centralized clouds. Advanced computing power, along with massive data acquisition, has enabled many time-sensitive operations, such as real-time monitoring and security analysis. However, with increasing complexity of power grids, such computing paradigm may face several challenges, such as privacy concern and communication bandwidth limit. In contrast, edge computing that leverages computing resources at edge has the potential to improve computation efficiency and protect data privacy by performing data analytic close to customers~\cite{feng2021smart}. Particularly, machine learning approaches that can preserve privacy, such as federated learning~\cite{mcmahan2017communication}, have drawn increasing attention.
\section{AI Solutions to Power Grid Decision Making}\label{sec:algorithm}
This section surveys recent AI solutions to the core decision making processes in power grid operations. We report $85$ papers, most of which were published in the IEEE transactions of the Power and Energy Society (e.g., IEEE Transactions on Power Systems, and IEEE Transactions on Smart Grid) from 2019 to 2021. For earlier works about AI algorithms for grid operations, we refer readers to previous survey papers\cite{LIU199286, 260875,duchesne2020recent,GLAVIC20176918}. Table \ref{tab:AI-classification} classifies the approaches used in these $85$ papers according to the category to which these approaches belong, (i.e., supervised, unsupervised, and reinforcement learning). In addition, for each decision making process, we provide not only an overview of the state-of-the-art, AI-powered grid solutions, but also illustrative examples that give readers a sense of how specific AI techniques can be leveraged to solve grid challenges. We use an independent notation system in each subsection,.
\begin{table}[tb]
	\caption{Classification of Grid Solutions based on AI Methods}
	\label{tab:AI-classification}
	\centering
\begin{tabular}{l|l|l|l}
	\hline

	\hline
	\textbf{\begin{tabular}[c]{@{}l@{}}Key Decision-\\making Module\end{tabular}} & \textbf{\begin{tabular}[c]{@{}l@{}}Supervised\\ Learning\end{tabular}} & \textbf{\begin{tabular}[c]{@{}l@{}}Unsupervised\\ Learning\end{tabular}} &\textbf{\begin{tabular}[c]{@{}l@{}}Reinforcement\\ Learning\end{tabular}} \\
	\hline
	     \begin{tabular}[c]{@{}l@{}} Renewable/Load\\ Modeling \end{tabular}& \begin{tabular}[c]{@{}l@{}}\cite{8946708,9066877,9334448,8291011,8080240,8068999,9350235,9335539,8854896,9031434,8616827,9447956,9531450,8743419,8580409,8735925,9371010,9601234,8854895}\end{tabular} &\begin{tabular}[c]{@{}l@{}} \cite{8080240,9335539}\\ \cite{9031434,8616827}\end{tabular}& \begin{tabular}[c]{@{}l@{}}\cite{8813103}\end{tabular} \\
	\hline
	     \begin{tabular}[c]{@{}l@{}} Grid Economic\\ Operation \end{tabular}& \begin{tabular}[c]{@{}l@{}}\cite{9454298,9115822,8802273,8999581,9599383,9388933,9335481,9205647,9524523,8017431,8424030,9034123,8960435,9426454,9483690,8957258,8336990}\end{tabular} &\begin{tabular}[c]{@{}l@{}}-\end{tabular}& \begin{tabular}[c]{@{}l@{}}-\end{tabular} \\
	     \hline
	     \begin{tabular}[c]{@{}l@{}} Grid Security\\ \& Resource Adequacy \end{tabular}& \begin{tabular}[c]{@{}l@{}} \cite{9606522,9234662,8418861,9146279,8649711,9477124,9524510,9165193,9606553,8892649,9055051,9445639,8839818}\end{tabular} &\begin{tabular}[c]{@{}l@{}} \cite{9559389,9606553}\end{tabular}& \begin{tabular}[c]{@{}l@{}}-\end{tabular} \\\hline
	     \begin{tabular}[c]{@{}l@{}}  Grid Monitoring,\\ Control, \& Protection \end{tabular}& \begin{tabular}[c]{@{}l@{}} \cite{8956077,8723612,9312447,8920121,8447254,9193985,9381611,9557843,8418852,8618342,9094728,8846225,9468376,8747534,9105102,8693907,8667699,8977490,8718345,8477148,9296965,9470951,8437176,9043691,8421644,9619930,9559412,9451602,9573436}\end{tabular} &\begin{tabular}[c]{@{}l@{}} \cite{8625444,9335975}\end{tabular}& \begin{tabular}[c]{@{}l@{}}\cite{8787888}\cite{9143169},\\ \cite{8356086}\end{tabular} \\

	\hline

	\hline
	\end{tabular}
\end{table}

\subsection{Renewable/Load Modeling}
Renewables and load introduce many uncertainties to the operation of low-carbon power grids. One way to address such uncertainties in grid operation is to develop an accurate forecast algorithm for renewables and load. The topic areas in renewable/load modeling include renewable (e.g., wind, and solar) generation forecasting, load forecasting, and load clustering. Table \ref{tab:load_modeling}\footnote{In the table ``-'' indicates that no computation resource is reported in reference.} lists the most recent works in these topic areas. Table \ref{tab:load_modeling} also summarizes the data source, AI method, and computation resource used in the references provided. 

\begin{table*}[tb]
	\caption{Tri-factors of AI-based Renewable/Load Forecasting}
	\label{tab:load_modeling}
	\centering

	\begin{tabular}{l|l|l|l}
	\hline
	
	\hline

	\textbf{Task [Ref.]} & \textbf{Data Sources {(Availability)}} &\textbf{Computation Resources}& \textbf{AI Method}  \\
	\hline
		 PV prediction on shipboards \cite{8946708}& Solar irradiance and weather data {(N)}&\begin{tabular}[c]{@{}l@{}}i7-7700 CPU\end{tabular}& \begin{tabular}[c]{@{}l@{}}Neural network (NN),\\ Extreme learning machine (ELM)\end{tabular}\\
		 \hline
		 Wind power prediction \cite{9066877} & \begin{tabular}[c]{@{}l@{}}Actual wind power data \\from Glens of Foudland wind farm {(N)}\end{tabular} & \begin{tabular}[c]{@{}l@{}}i7-7700 CPU, 16GB RAM \end{tabular}& \begin{tabular}[c]{@{}l@{}}ELM\end{tabular} \\
		 \hline
		 Wind power prediction \cite{9334448}&
		 \begin{tabular}[c]{@{}l@{}}Wind generation data from\\ Irish transmission system operator {(N)}\end{tabular} & \begin{tabular}[c]{@{}l@{}}i7-7700 CPU; 16GB RAM  \end{tabular}& ELM \\
		 \hline
		 \begin{tabular}[c]{@{}l@{}}Socio-demographic \\information identification\cite{8291011}\end{tabular}&Residential smart meter data {(N)}& \begin{tabular}[c]{@{}l@{}}i7-4770MQ CPU\end{tabular} & Convolutional neural network (CNN)  \\
		 \hline
		 \begin{tabular}[c]{@{}l@{}}Feature selection \\for PV forecasting \cite{ 8080240}\end{tabular}&\begin{tabular}[c]{@{}l@{}}Three PV arrays' measured \\datasets in Australia {(N)}\end{tabular} & -&\begin{tabular}[c]{@{}l@{}} Principal component analysis (PCA);\\ K-nearest neighbors (KNN)\end{tabular} \\
		 \hline
		 \begin{tabular}[c]{@{}l@{}}Wind power\\ramp forecasting\cite{8068999}\end{tabular}&\begin{tabular}[c]{@{}l@{}}Wind Integration National\\Dataset (WIND) Toolkit \cite{8068999} {(Y)}\end{tabular}&\begin{tabular}[c]{@{}l@{}} Intel-e5-2603 32 GB RAM\end{tabular}&Gaussian mixture model\\
		 \hline
         Solar forecasting \cite{9350235} & \begin{tabular}[c]{@{}l@{}}Desert Knowledge Australia \\Solar Centre (DKASC) {(N)}\end{tabular}&
	    Nvidia GTX 1080 GPU &  Spectral graph convolution \\
	     \hline
	     \begin{tabular}[c]{@{}l@{}} PV forecasting \cite{9335539}\end{tabular}& \begin{tabular}[c]{@{}l@{}} DKASC  {(N)}\end{tabular} &-& \begin{tabular}[c]{@{}l@{}}ELM, Auto-encoder \end{tabular} \\
         \hline
	     \begin{tabular}[c]{@{}l@{}} Short-term load forecasting \cite{ 8854896}\end{tabular}& \begin{tabular}[c]{@{}l@{}} Actual data of Tianjin\\power Grid in China  {(N)}\end{tabular} &\begin{tabular}[c]{@{}l@{}} i5-6700 CPU, 16GB RAM\end{tabular}& \begin{tabular}[c]{@{}l@{}} Deep belief network\end{tabular} \\
	     
	     \hline
	     \begin{tabular}[c]{@{}l@{}}Load forecasting \cite{8813103}\end{tabular}& \begin{tabular}[c]{@{}l@{}} Hourly load data of \\the University of Texas\\ at Dallas (UTD) {(Y)}\end{tabular} &\begin{tabular}[c]{@{}l@{}} i7-4870HQ CPU, 16GB RAM \end{tabular}& \begin{tabular}[c]{@{}l@{}} Reinforcement learning\end{tabular} \\
	     
	     \hline
	     \begin{tabular}[c]{@{}l@{}}Customer segmentation \cite{9031434 }\end{tabular}& \begin{tabular}[c]{@{}l@{}} Real-world utility data {(N)}\end{tabular} &-& \begin{tabular}[c]{@{}l@{}} Spectral clustering; regression
\end{tabular} \\
\hline
	     \begin{tabular}[c]{@{}l@{}}Load clustering, \\prediction, and inference \cite{ 8616827 }\end{tabular}& \begin{tabular}[c]{@{}l@{}}18-node real utility feeder {(N)}\end{tabular} &-& \begin{tabular}[c]{@{}l@{}} Spectral clustering;\\ Recursive Baysian Learning\end{tabular} \\

	     \hline
	     \begin{tabular}[c]{@{}l@{}}Wind power prediction \cite{9447956}\end{tabular}& \begin{tabular}[c]{@{}l@{}}Data from three wind farm \\data in China {(N)}\end{tabular} &-& \begin{tabular}[c]{@{}l@{}} Multi-source and temporal\\ attention network \end{tabular} \\
	     
	     \hline
	     \begin{tabular}[c]{@{}l@{}}Demand flexibility estimation
\cite{9531450}
\end{tabular}& \begin{tabular}[c]{@{}l@{}}Datasets from ERCOT, \\PJM, CAISO {(Y)}\end{tabular} &\begin{tabular}[c]{@{}l@{}} i7-8550U CPU, 16GB RAM \end{tabular}& \begin{tabular}[c]{@{}l@{}} Long short term memory (LSTM) NNs \end{tabular} \\

	     \hline
	     \begin{tabular}[c]{@{}l@{}}Wind power prediction \cite{8743419}
\end{tabular}& \begin{tabular}[c]{@{}l@{}}Wind farm data from NREL {(Y)}\end{tabular} &\begin{tabular}[c]{@{}l@{}} i7-7700 CPU, 16GB RAM  \end{tabular}& \begin{tabular}[c]{@{}l@{}} ELM \end{tabular} \\

	     \hline
	     \begin{tabular}[c]{@{}l@{}}PV power prediction \cite{8580409 }
\end{tabular}& \begin{tabular}[c]{@{}l@{}}5-MW PV power plant {(Y)} \end{tabular} &\begin{tabular}[c]{@{}l@{}}i7-2600 CPU\end{tabular}& \begin{tabular}[c]{@{}l@{}} Deep belief network  \end{tabular} \\
\hline
	     \begin{tabular}[c]{@{}l@{}}Cold load pick-up \\demand
assessment \cite{ 8735925 }
\end{tabular}& \begin{tabular}[c]{@{}l@{}}Real-world smart meter data {(N)}\end{tabular}&\begin{tabular}[c]{@{}l@{}}-\end{tabular} & \begin{tabular}[c]{@{}l@{}} Regression; Gaussian mixture model\end{tabular} \\

\hline
	     \begin{tabular}[c]{@{}l@{}}Dynamic load modeling \cite{ 9371010 }
\end{tabular}& \begin{tabular}[c]{@{}l@{}}CIGRE benchmark \\low voltage network {(N)}\end{tabular} &\begin{tabular}[c]{@{}l@{}}-\end{tabular}& \begin{tabular}[c]{@{}l@{}} Decision trees;\\
Ant colony optimization
\end{tabular} \\

\hline
	     \begin{tabular}[c]{@{}l@{}}PV power forecasting \cite{9601234 }
\end{tabular}& \begin{tabular}[c]{@{}l@{}}PV plant dataset {(N)}\end{tabular}&\begin{tabular}[c]{@{}l@{}}-\end{tabular} & \begin{tabular}[c]{@{}l@{}} Graph neural network
\end{tabular} \\

\hline
	     \begin{tabular}[c]{@{}l@{}}Load forecasting \cite{8854895}
\end{tabular}& \begin{tabular}[c]{@{}l@{}}Real data set from \\residential Irish customers {(N)}\end{tabular} &\begin{tabular}[c]{@{}l@{}}-\end{tabular}& \begin{tabular}[c]{@{}l@{}} Random forest
\end{tabular} \\

	\hline

	\hline
	\end{tabular}
\end{table*}

\begin{figure}
    \centering
    \includegraphics[width = 3in]{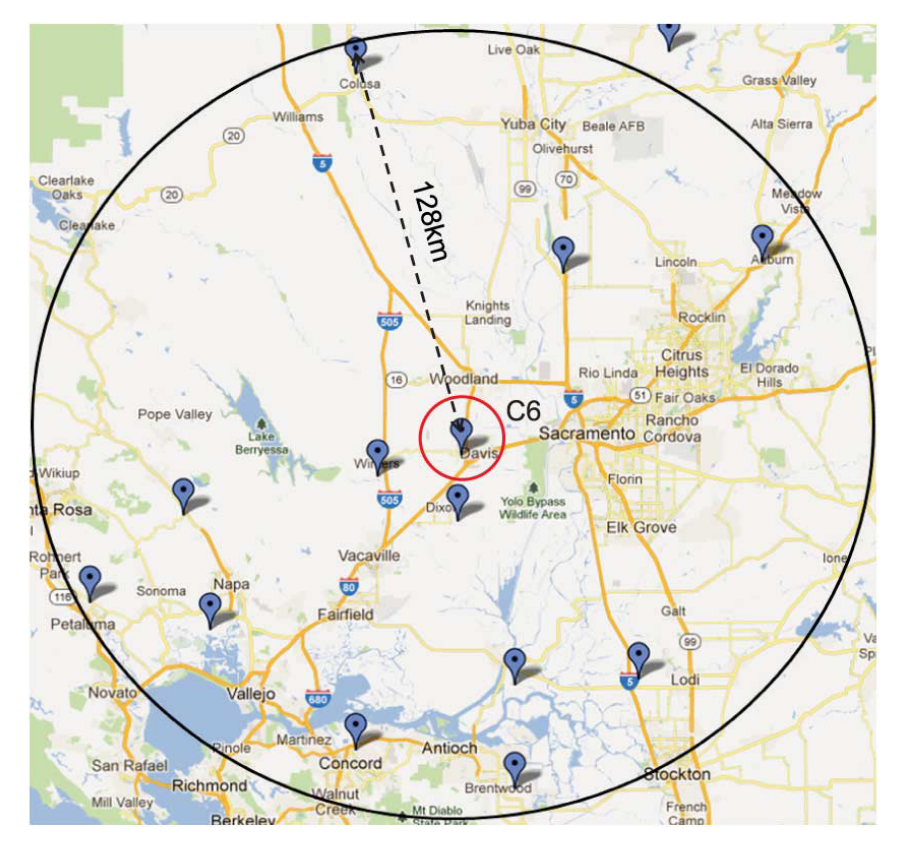}
    \caption{The target solar farm site C6 (in the red circle) and its neighboring solar farms. (Source: Fig. 1 of \cite{6945846} \copyright IEEE 2015)}
    \label{fig:PV_forecast}
\end{figure}

Next we provide an example to elaborate on how AI can be leveraged to solve PV forecasting tasks in the grid. The technical details are reported in \cite{6945846}. Figure \ref{fig:PV_forecast} shows the geographic locations of a target solar site C6 and its neighboring $N$ solar sites. Let us suppose that we want to predict the solar irradiance of the target solar site C6 at time step $(k+1)$. Reference \cite{6945846} formulates the forecasting problem into one of estimating the parameters of the following autoregressive with exogenous input (ARX) model\cite{6945846}:
\begin{equation} \label{eq:ARX_form}
    \begin{aligned}
     x[k+1] = f(x[k], &\ldots, x[k-n+1],\\
              w_1[k-d_1+1], &\ldots, w_1[k-m_1-d_1 +2],\ldots\\
              w_i[k-d_i+1], &\ldots, w_i[k-m_i-d_i + 2],\ldots\\
              w_N[k-d_N+1], &\ldots, w_N[k-m_N-d_N +2])
    \end{aligned}
\end{equation}
where $x[k]$ is the solar irradiance at the target solar site at time step $k$; $w_i$ is the solar irradiance at the neighboring solar site $i$; $f(\cdot)$ is an ARX-structured function; and positive integers $n$, $d_i$, and $m_i$ are user-defined parameters that can be determined at training stages \cite{6945846}. The intuition of the formulation \eqref{eq:ARX_form} is that the next-step solar irradiance $x[k+1]$ at the target solar site depends not only on the local solar irradiance, but also on the solar irradiance at its neighboring solar sites. The case studies based on real-world renewable data from California and Colorado suggest such an algorithm is suitable for 1-h and 2-h ahead PV forecasting \cite{6945846}. However, the algorithm proposed in \cite{6945846} dose not provide a probability description for the forecast quality. One potential avenue for future work is to investigate such a description \cite{6945846}.

\subsection{Grid Economic Operation}
The large-scale deployment of renewables poses unprecedented challenges to the electricity market operation. Conventional deterministic tools may not be able to support the electricity market operation of the electricity infrastructure with a significant amount of uncertain renewables. Reference \cite{8485380} proposes a scenario-based approach that unlocks the potential of data in order to incorporate renewables' uncertainties into the dispatch of grid resources. Let us suppose that there are $N$ historical scenarios $\Delta_N=\{\delta_1, \delta_2, \ldots, \delta_n, \ldots, \delta_{N}\}$ that is a subset of all possible scenarios $\Delta$. In each historic scenario $\delta_n$, the net-load forecasting errors at each bus is recorded. Reference \cite{8485380} formulates the ED problem as follows \cite{8485380}:
\begin{subequations} \label{eq:Sc_Formulation}
    \begin{align}
    \min_{p} \quad& c^{\top}p \\
    \text{s.t.} \quad& g_1(p) \le 0 \label{eq:scenario_independent}\\
    & g_2(p, \delta_n) \le 0, \forall \delta_n \in \Delta_N\label{eq:scenario_dependent}
    \end{align}
\end{subequations}
where vector $p$ concerns the power generation of all generators in all intervals during a planning horizon; vector $c$ collects cost coefficients associated with generators; \eqref{eq:scenario_independent} represents the scenario-independent constraints \cite{8485380}, such as ramp and capacity constraints of generators; and \eqref{eq:scenario_dependent} represents the scenario-dependent constraints \cite{8485380}, such as generation-load balance constraints. Suppose that $p^*_N$ is the solution to the optimization \eqref{eq:Sc_Formulation} given $N$ historical samples. Because $\Delta_N$ is a subset of all possible scenarios $\Delta$, it is possible that there exists a scenario $\delta$ that causes the scenario-dependent constraints to be violated, i.e., $g_2(p^*_N, \delta)>0$. The probability that such an event may occur is termed the ``risk" in \cite{8485380}. Formally, the risk $v(p^*_N)$ for the solution $p^*_N$ is defined by
\begin{equation}
    v(p^*_N) = \text{Prob.}(\delta \in \Delta: g_2(p^*_N, \delta)>0)
\end{equation}
where $\text{Prob.}(\cdot)$ denotes the probability that event ``$\cdot$'' occurs. 
We expect that the probability that the risk $v(p^*_N)$ of solution $p^*_N$ exceeds a small number $\epsilon$ will be small, i.e.,
\begin{equation} \label{eq:risk_preference}
    \text{Prob.}(v(p^*_N)>\epsilon)<\gamma
\end{equation}
where $0<\epsilon, \gamma\ll1$. With the risk preference parameters $\epsilon$ and $\gamma$, a natural question is how to determine the size of $\Delta_N$, i.e., $N$, to achieve the risk preference \eqref{eq:risk_preference}. Reference \cite{8485380} provides a lower bound of $N$ that depends solely on the look-ahead intervals and risk preference parameters \cite{8485380}. Such a lower bound can help system operators determine how many scenarios must be drawn from the historical observations based on their risk preference. For example, in an open-source, 2000-bus synthetic Texas grid, if we suppose that the risk preference parameters of the system operators are $\gamma = 10^{-6}$ and $\epsilon = 0.0083$, then 2000 historical scenarios are needed to be embedded into the ED formulation \eqref{eq:Sc_Formulation} \cite{8485380}. A rigorous investigation of the relationship between the quantity of support constraints and the design parameters ($\gamma$ and $\epsilon$) is still needed to further refine the algorithm in \cite{8485380}.

Other recent AI solutions to the problems of UC, ED, and OPF are summarized in Table \ref{tab:grid_economic}. The AI methods associated with the data sources and computation resources in the references are listed in Table \ref{tab:grid_economic}.
\begin{table*}[tb]
	\caption{Tri-factors of AI Solutions to Market Operation}
	\label{tab:grid_economic}
	\centering

	\begin{tabular}{l|l|l|l}
	\hline

	\hline
	\textbf{Task [Ref.]} & \textbf{Data Resources {(Availability)}} & \textbf{Computation Resources} &\textbf{AI Methods} \\
	\hline
	     \begin{tabular}[c]{@{}l@{}}UC \& ED \cite{ 9454298} \end{tabular}& \begin{tabular}[c]{@{}l@{}}A real-world 1881-bus\\ system in China {(N)}\end{tabular} &\begin{tabular}[c]{@{}l@{}} i7-1065G7 CPU, 16GB RAM\end{tabular}& \begin{tabular}[c]{@{}l@{}}NN and oblique \\decision trees  \end{tabular} \\
	\hline
	     \begin{tabular}[c]{@{}l@{}}OPF \cite{ 9115822} \end{tabular}& \begin{tabular}[c]{@{}l@{}}Polish 2383-bus system  {(N)}\end{tabular}&\begin{tabular}[c]{@{}l@{}}i7-8700K CPU, 32GB RAM\end{tabular} & \begin{tabular}[c]{@{}l@{}}Stacked ELM\end{tabular} \\
	     
	\hline
	     \begin{tabular}[c]{@{}l@{}}OPF \cite{ 8802273}\end{tabular}& \begin{tabular}[c]{@{}l@{}}129-node feeder {(N)}\end{tabular}&\begin{tabular}[c]{@{}l@{}}- \end{tabular} & \begin{tabular}[c]{@{}l@{}}Regression\end{tabular} \\
		\hline
	     \begin{tabular}[c]{@{}l@{}}Probabilistic power flow \cite{8999581}\end{tabular}& \begin{tabular}[c]{@{}l@{}}A 33 bus distribution \\system from MATPOWER\end{tabular} &\begin{tabular}[c]{@{}l@{}}i7-7820HQ CPU, 31.9GB RAM\end{tabular}& \begin{tabular}[c]{@{}l@{}}Copula function\end{tabular} \\
	     \hline
	     \begin{tabular}[c]{@{}l@{}}ACOPF\cite{9599383}\end{tabular}& \begin{tabular}[c]{@{}l@{}}A 6700-bus French system {(N)};\\ a 9000-bus system {(N)}\end{tabular}&\begin{tabular}[c]{@{}l@{}}i7 CPU 16GB RAM;\\Nvidia Tesla V100 GPUs, 16GB RAM
\end{tabular} & \begin{tabular}[c]{@{}l@{}}NN \end{tabular} \\
\hline
	     \begin{tabular}[c]{@{}l@{}}Binding constrain \\prediction in UC \cite{9388933}\end{tabular}& \begin{tabular}[c]{@{}l@{}}500-bus synthetic \\South Carolina system {(N)}\end{tabular} &\begin{tabular}[c]{@{}l@{}} 24-core CPU, 128GB RAM\end{tabular}& \begin{tabular}[c]{@{}l@{}}Bagged trees\end{tabular} \\
	     \hline
	     \begin{tabular}[c]{@{}l@{}}SCOPF\cite{9335481}\end{tabular}& \begin{tabular}[c]{@{}l@{}}118-IEEE, 1354-PEG;\\ 1888-RTE system {(N)}\end{tabular} &\begin{tabular}[c]{@{}l@{}} Nvidia Tesla V100 GPUs\end{tabular}& \begin{tabular}[c]{@{}l@{}}Deep learning\end{tabular} \\
	     \hline
	     \begin{tabular}[c]{@{}l@{}}OPF\cite{9205647}\end{tabular}& \begin{tabular}[c]{@{}l@{}}30-, 57-, 118-, 300-IEEE {(N)}\end{tabular} &\begin{tabular}[c]{@{}l@{}} 4-core i7-3770 CPU, 16GB RAM \end{tabular}& \begin{tabular}[c]{@{}l@{}}Deep learning\end{tabular} \\
	     \hline
	     \begin{tabular}[c]{@{}l@{}}Financial return\\maximization \cite{ 9524523}\end{tabular}& \begin{tabular}[c]{@{}l@{}}261,489-consumer real dataset\\ from a Brazilian utility {(N)}\end{tabular} &\begin{tabular}[c]{@{}l@{}}-\end{tabular}& \begin{tabular}[c]{@{}l@{}}Rotation forest;\\ XGBoost \end{tabular} \\
	      \hline
	     \begin{tabular}[c]{@{}l@{}}Energy pricing \cite{8017431}
\end{tabular}& \begin{tabular}[c]{@{}l@{}}13-IEEE {(N)}\end{tabular} &-& \begin{tabular}[c]{@{}l@{}} Regression \end{tabular} \\

	      \hline
	     \begin{tabular}[c]{@{}l@{}}Energy bidding \cite{ 8424030 }
\end{tabular}& \begin{tabular}[c]{@{}l@{}}Real-world price data {(Y)}\end{tabular}&- & \begin{tabular}[c]{@{}l@{}} Risk-averse learning \end{tabular} \\
	     \hline
	     \begin{tabular}[c]{@{}l@{}}Constrain screening of UC
\cite{ 9034123 }
\end{tabular}& \begin{tabular}[c]{@{}l@{}}IEEE RTS-96;\\2000-bus Texas synthetic grid {(N)}\end{tabular}&\begin{tabular}[c]{@{}l@{}}2.6 GHz CPU, 2GB RAM\end{tabular} & \begin{tabular}[c]{@{}l@{}}KNN\\ XGBoost \end{tabular} \\

  \hline
	     \begin{tabular}[c]{@{}l@{}}Reserve capacity \\estimation \cite{ 8960435 }
\end{tabular}& \begin{tabular}[c]{@{}l@{}}821-day data {(N)}\end{tabular}&\begin{tabular}[c]{@{}l@{}}-\end{tabular} & \begin{tabular}[c]{@{}l@{}}ELM \end{tabular} \\

\hline
	     \begin{tabular}[c]{@{}l@{}}ED and UC \cite{9426454 }
\end{tabular}& \begin{tabular}[c]{@{}l@{}}24-, 118- IEEE {(N)}\end{tabular}&\begin{tabular}[c]{@{}l@{}}Intel Xeon CPU, 16GB RAM \end{tabular} & \begin{tabular}[c]{@{}l@{}}XGBoost  \end{tabular} \\

\hline
	     \begin{tabular}[c]{@{}l@{}}Energy bidding \cite{9483690 }
\end{tabular}& \begin{tabular}[c]{@{}l@{}}Market data from PJM, \\CAISO, and ISO-NE {(N)}\end{tabular}&\begin{tabular}[c]{@{}l@{}}- \end{tabular} & \begin{tabular}[c]{@{}l@{}}XGBoost \end{tabular} \\
\hline
	     \begin{tabular}[c]{@{}l@{}}Electricity price scenario \\generation \cite{8957258 }
\end{tabular}& \begin{tabular}[c]{@{}l@{}}Dutch market prices {(N)}\end{tabular}&\begin{tabular}[c]{@{}l@{}}Nvidia Tesla V100, 61GB RAM \end{tabular} & \begin{tabular}[c]{@{}l@{}}LSTM; recurrent\\ neural network \end{tabular} \\

\hline
	     \begin{tabular}[c]{@{}l@{}}Price forecasting \cite{ 8336990}
\end{tabular}& \begin{tabular}[c]{@{}l@{}}Historic price data {(N)}\end{tabular}&\begin{tabular}[c]{@{}l@{}}- \end{tabular} & \begin{tabular}[c]{@{}l@{}}ELM \end{tabular} \\

	\hline

	\hline
	\end{tabular}
\end{table*}


\subsection{Grid Security and Resource Adequacy}
To decarbonize the power grids, fossil-fueled generators are being replaced by inverter-based resources, e.g., wind/solar farms, and energy storage. To assess grid security and resource adequacy, it is necessary to develop new planning tools that explicitly consider these new elements. The grid security and resource adequacy analysis include steady-state,dynamic security analysis, and reliability analyses. Table \ref{tab:grid_security} summarizes the state-of-the-art AI adoption in these analyses. Next, we will present a learning-based approach to networked microgrid security analysis \cite{9559389}, in order to show how an AI technique can be adopted in this specific topic area. 

\begin{table*}[tb]
	\caption{Tri-factors of AI Solutions to Grid Security and Resource Adequacy Analysis}
	\label{tab:grid_security}
	\centering

	\begin{tabular}{l|l|l|l}
	\hline

	\hline
	\textbf{Task [Ref.]} & \textbf{Data Sources {(Availability)}} &\textbf{Computation Resources}& \textbf{AI Method}  \\
	\hline
	     \begin{tabular}[c]{@{}l@{}}Reliability study for \\power-gas systems \cite{9606522}\end{tabular}& \begin{tabular}[c]{@{}l@{}}73-bus power system \\w/ 40-node gas system {(N)}\end{tabular} &-& \begin{tabular}[c]{@{}l@{}}Random Forest; 
XGBoost\end{tabular} \\
\hline
	     \begin{tabular}[c]{@{}l@{}}Reliability study \\w/ rich PE \cite{ 9234662}\end{tabular}& \begin{tabular}[c]{@{}l@{}}IEEE RTS-24 bus \\w/ 40-node gas system {(N)}\end{tabular} &\begin{tabular}[c]{@{}l@{}} Intel CPU, 16GB RAM \end{tabular}& \begin{tabular}[c]{@{}l@{}}Support vector regression \\(SVR); random forests (RF) \end{tabular} \\
	     \hline
	     \begin{tabular}[c]{@{}l@{}}Energy Loss estimation \cite{ 8418861}\end{tabular}& \begin{tabular}[c]{@{}l@{}}33-bus distribution system \\w/ 40-node gas system {(N)}\end{tabular}&\begin{tabular}[c]{@{}l@{}}PC, 8GB RAM\end{tabular} & \begin{tabular}[c]{@{}l@{}}Regression trees \end{tabular} \\
	    \hline
	     \begin{tabular}[c]{@{}l@{}}Distribution system\\phase identification \cite{9146279}\end{tabular}& \begin{tabular}[c]{@{}l@{}}25-bus, 123-bus, \\450-bus systems {(N)}\end{tabular} &-& \begin{tabular}[c]{@{}l@{}}Modified k-means \end{tabular} \\
	     \hline
	     \begin{tabular}[c]{@{}l@{}}Outage scheduling \cite{ 8649711}\end{tabular}& \begin{tabular}[c]{@{}l@{}}IEEE-RTS79;\\IEEE- RTS96 {(N)}\end{tabular}&\begin{tabular}[c]{@{}l@{}}300-core Xeon CPUs, \\2GB RAM for each\end{tabular} & \begin{tabular}[c]{@{}l@{}}k-nearest neighbor \\classification  \end{tabular}  \\
	\hline
	     \begin{tabular}[c]{@{}l@{}}Energy Disaggregation \\at Substations \cite{ 9477124 }\end{tabular}& \begin{tabular}[c]{@{}l@{}}Ind-Solar dataset;\\ 
EnerNOC GreenButton Data;\\
Solar generation from National\\ Renewable Energy Laboratory {(N)}\end{tabular} &\begin{tabular}[c]{@{}l@{}}i9 CPU, 32 GB RAM\end{tabular}& \begin{tabular}[c]{@{}l@{}}Bayesian dictionary learning\end{tabular}  \\
 
 \hline
	     \begin{tabular}[c]{@{}l@{}}Distribution power flow \cite{9524510}\end{tabular}& \begin{tabular}[c]{@{}l@{}}8-, 123- IEEE; 362-node\\ utility distribution network
 {(N)}\end{tabular} &-& \begin{tabular}[c]{@{}l@{}}Support Matrix Regression\end{tabular} \\
 
 \hline
	     \begin{tabular}[c]{@{}l@{}}Microgrid Scheduling \cite{9165193}\end{tabular}& \begin{tabular}[c]{@{}l@{}}33-node system
 {(Y)}\end{tabular}&- & \begin{tabular}[c]{@{}l@{}}Deep learning\end{tabular} \\
 
 \hline
	     \begin{tabular}[c]{@{}l@{}}Stability Assessment \\of Networked Microgrids \cite{ 9559389}\end{tabular}& \begin{tabular}[c]{@{}l@{}}123-IEEE
 {(N)}\end{tabular} &\begin{tabular}[c]{@{}l@{}}i5 CPU, 8GB RAM \end{tabular}& \begin{tabular}[c]{@{}l@{}}Neural Lyapunov method\end{tabular}  \\
 
 \hline
	     \begin{tabular}[c]{@{}l@{}}Residential PV localization \cite{ 9606553}\end{tabular}& \begin{tabular}[c]{@{}l@{}}Umass Smart data 
 {(Y)}\end{tabular} &\begin{tabular}[c]{@{}l@{}}Xeon CPU E5-2687W v4,\\64 GB RAM  \end{tabular}& \begin{tabular}[c]{@{}l@{}}Autoencoder; semi-supervised \\learning\end{tabular}  \\
 
 \hline
	     \begin{tabular}[c]{@{}l@{}}Phase Identification
\cite{ 8892649 }
\end{tabular}& \begin{tabular}[c]{@{}l@{}}Systems from SCE, \\PGEC, and FortisBC
 {(Y)}\end{tabular} &\begin{tabular}[c]{@{}l@{}}-\end{tabular}& \begin{tabular}[c]{@{}l@{}}Information theoretic machine \\learning \end{tabular}  \\
 
 \hline
	     \begin{tabular}[c]{@{}l@{}}Reliability study \cite{9055051}
\end{tabular}& \begin{tabular}[c]{@{}l@{}}ISO-NE area data 
 {(Y)}\end{tabular}&\begin{tabular}[c]{@{}l@{}}-\end{tabular} & \begin{tabular}[c]{@{}l@{}}Regression\end{tabular}  \\
 
 \hline
	     \begin{tabular}[c]{@{}l@{}}Extreme outage prediction \cite{9445639 }
\end{tabular}& \begin{tabular}[c]{@{}l@{}}24-IEEE
 {(N)}\end{tabular} &\begin{tabular}[c]{@{}l@{}}-\end{tabular}& \begin{tabular}[c]{@{}l@{}}Bayes Decision Theory\end{tabular}  \\
 
 \hline
	     \begin{tabular}[c]{@{}l@{}}Network management \cite{8839818 }
\end{tabular}& \begin{tabular}[c]{@{}l@{}}4,000 circuits from\\ U.K. utilities
 {(N)}\end{tabular} &\begin{tabular}[c]{@{}l@{}}-\end{tabular}& \begin{tabular}[c]{@{}l@{}}Decision tree\end{tabular}  \\
 
	\hline

	\hline
	\end{tabular}
\end{table*}

\begin{figure}[thb]
    \centering
    \includegraphics[width = 3.5in]{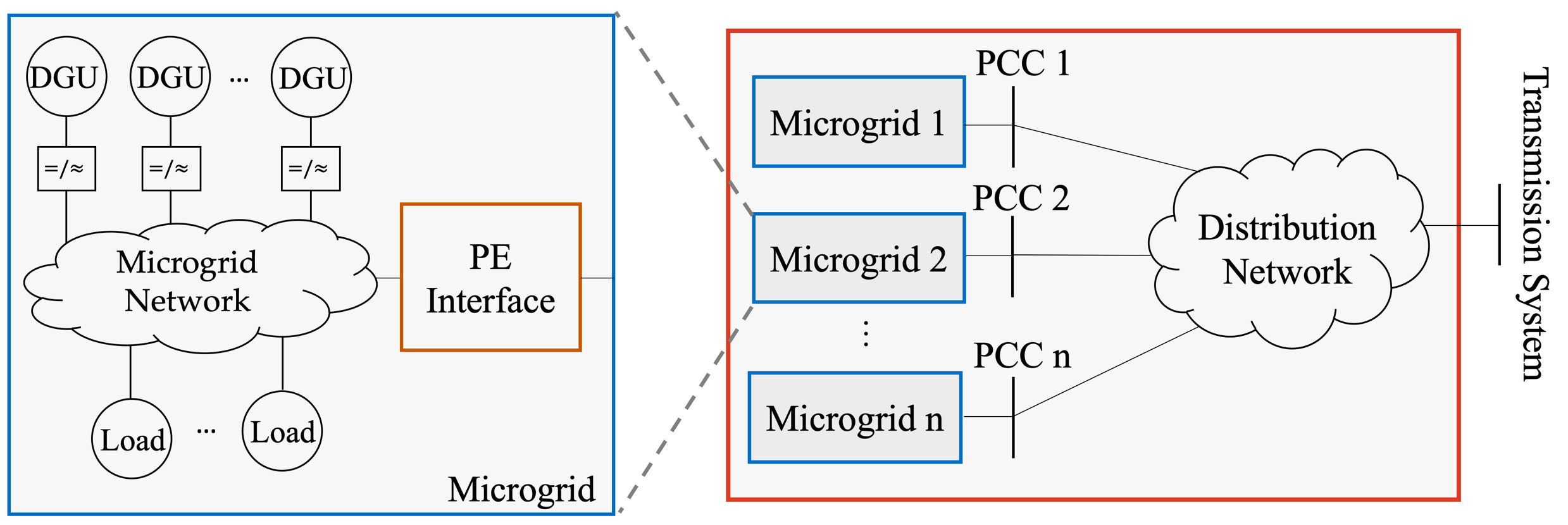}
    \caption{Physical architecture of networked microgrids with power electronics interfaces. (Source: Fig. 1 of reference \cite{9559389} \copyright IEEE 2021)}
    \label{fig:networked_MG}
\end{figure}

Fig. \ref{fig:networked_MG} shows the physical architecture of $n$ networked microgrids where the $n$ microgrids interact with one another via distribution lines. The dynamics of the networked microgrids can be described by $\dot{\mathbf{x}}= f (\mathbf{x})$, where state vector $\mathbf{x}$ is related to voltage magnitudes and phase angles at the points of common coupling (PCCs). In the networked microgrids, large disturbances may come from (i) the microgrid operating mode change, e.g., one microgrid enters an islanded mode; and (ii) the distribution network, e.g, distribution line tripping. The security analysis attempts to quantify the disturbance magnitude that the networked microgrids can tolerate \cite{9559389}. The result of this analysis is critical for both distribution system planners and operators.

\begin{figure}[thb]
    \centering
    \includegraphics[width = 3in]{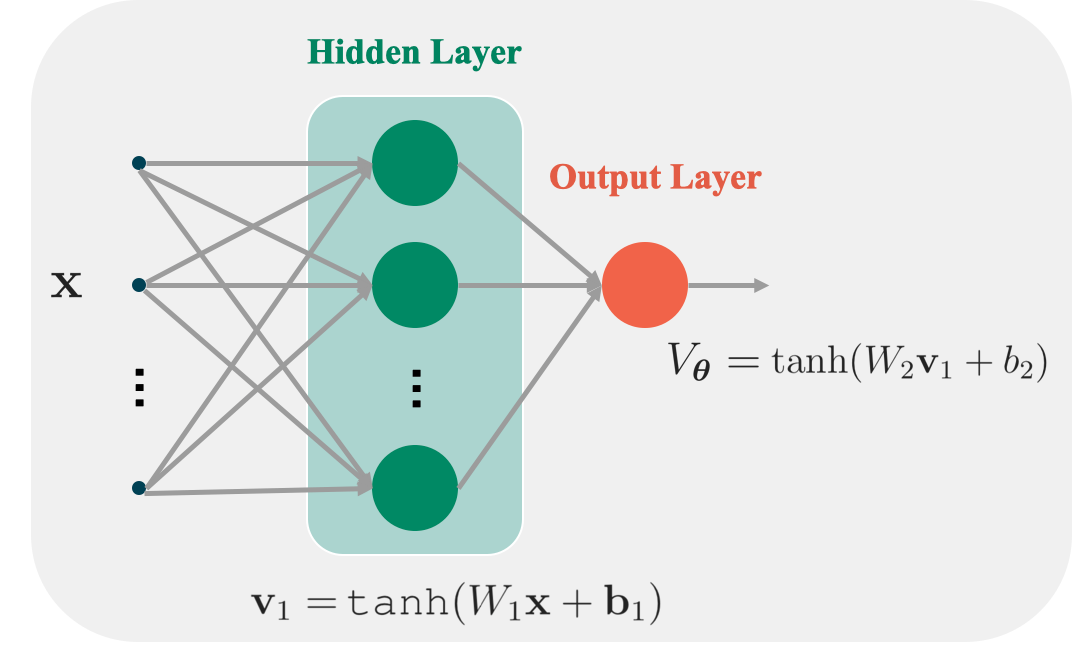}
    \caption{A neural network-structured Lyapunov function: the tunable parameter vector $\boldsymbol{\theta}$ is related to weights $W_1$ and $W_2$ and biases $\mathbf{b}_1$ and $b_2$ in the hidden and output layers. (Source: Fig. 3 of reference \cite{9559389} \copyright IEEE 2021)}
    \label{fig:NN_Lyapunov_function}
\end{figure}

\begin{figure}
    \centering
    \includegraphics[width = 2.5in]{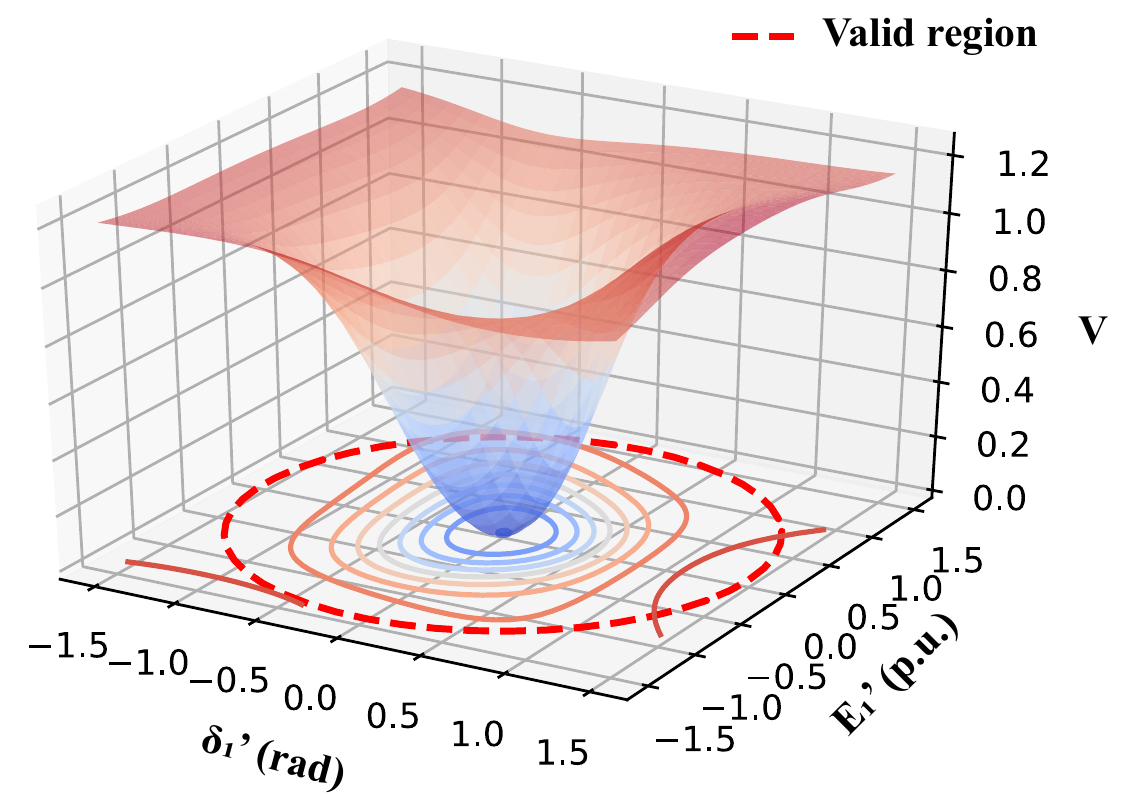}
    \caption{A Lyapunov function learned for a grid-tied microgrid. (Source: Fig. 6-a of reference \cite{9559389} \copyright IEEE 2021)}
    \label{fig:Lyapunov_function}
\end{figure}

\begin{figure}
    \centering
    \includegraphics[width = 2.5in]{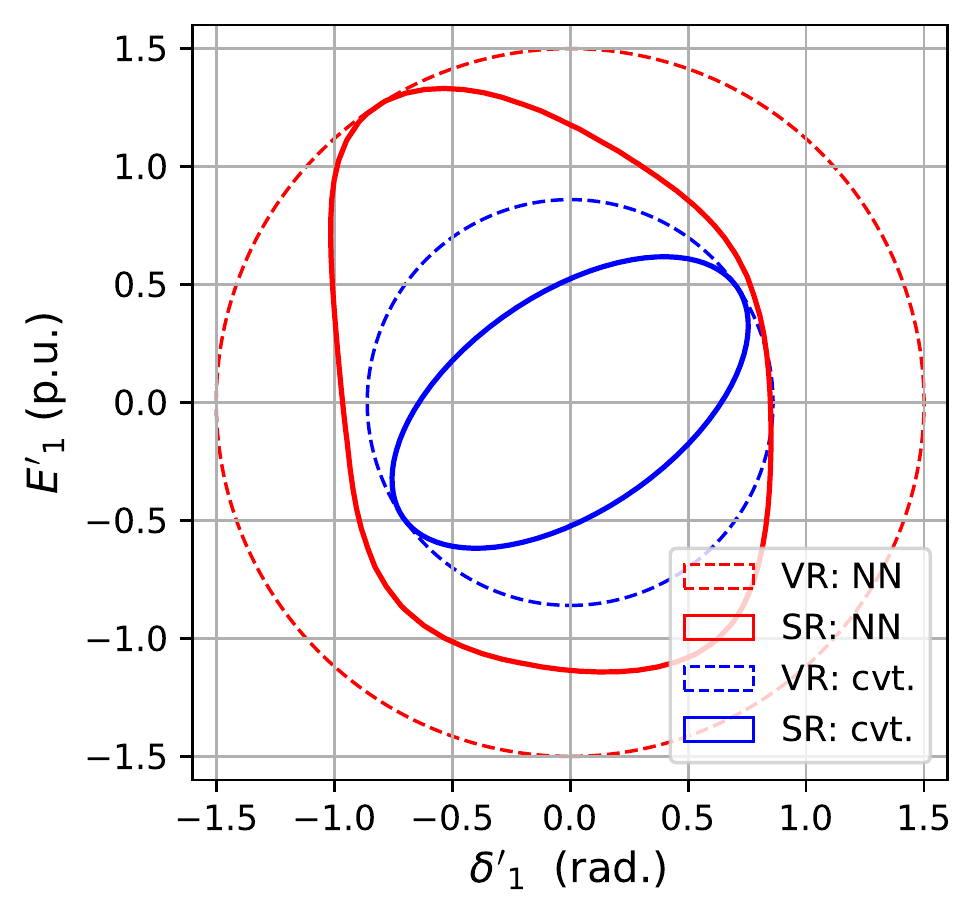}
    \caption{Security regions (SR) and valid regions (VR): the neural network (NN) approach and the conventional (cvt.) approach (Source: Fig. 7-a of reference \cite{9559389} \copyright IEEE 2021)}
    \label{fig:SR_comparison}
\end{figure}

In \cite{9559389}, Huang \emph{et al.} formulate the security analysis problem as one of searching for a legitimate Lyapunov function, i.e., a system-behavior summary function for a dynamic system. A Lyapunov function $V(\mathbf{x})$ satisfies two conditions: (i) $V(\mathbf{x})$ is a positive-definite function in a region $\mathcal{R}$ around the system equilibrium point; and (ii) the time derivative $\dot{V}$ is a negative-definite function in $\mathcal{R}$. In \cite{9559389}, the Lyapunov function is assumed to possess a neural network (NN) structure with parameter vector $\boldsymbol{\theta}$. To make the NN-structured function satisfy the two conditions of a Lyapunov function, a cost function $c(\boldsymbol{\theta})$ is designed. The cost function incurs a positive penalty if the NN with $\boldsymbol{\theta}$ violates one or both of the two Lyapunov function conditions. Vector $\boldsymbol{\theta}$ is tuned by the following procedure:
\begin{enumerate}
    \item Create a sample pool by randomly drawing a large number of states $\mathbf{x}$ within the region $\mathcal{R}$;
    \item Update $\boldsymbol{\theta}$ $n$ times based on the cost function $c(\boldsymbol{\theta})$ and the gradient descent algorithm \cite{9559389};
    \item For the NN with the latest $\boldsymbol{\theta}$, search for samples that violated one or both of the two Lyapunov conditions via the satisfiability modulo theories (SMT) tool. If no sample is found, claim the NN is a Lyapunov function; otherwise, add the samples to the sample pool in step 1) and repeat step 2).
\end{enumerate}
Fig. \ref{fig:NN_Lyapunov_function} visualizes a Lyapunov function learned from a state space for a grid-tied microgrid \cite{9559389}. The parameters of the system are reported in \cite{9559389}. It take $32.18$ seconds to learn the Lyapunov function \cite{9559389}. Having learned the Lyapunov function shown in Fig.~\ref{fig:Lyapunov_function}, a security region can be estimated, which is visualized in Fig. \ref{fig:SR_comparison}. If a disturbance leads the state vector to deviate from the equilibrium (the origin of Fig. \ref{fig:SR_comparison}) while also remaining within the solid red circle in Fig. \ref{fig:SR_comparison}, one can conclude immediately that the system trajectory will converge to the equilibrium without conducting any simulations. The region in the solid blue circle is the security region estimated by a conventional approach. It can be observed in Fig. \ref{fig:SR_comparison} that the learning-based approach is much less conservative than the conventional approach, since the red-solid circle is larger than the blue circle. Although the approach in \cite{9559389} can address heterogeneous interface dynamics and can provide less conservative results than the conventional approach, it incurs large computational costs when analyzing large-scale systems.

\subsection{Grid Monitoring, Control, and Protection}
Deep penetration of clean energy resources is changing power grid behaviour (for example, clean-energy resources may lack physical inertia). As a result, the power grids are becoming increasingly sensitive to disturbances and impact anomalies may become more frequently observable. Effectively monitoring and correcting these anomalies in real-time defines a key challenge facing system operators. A large body of literature in the last three years has argued in favor of leveraging streaming data to make operational decisions in real time. Table \ref{tab:grid_online} summarizes these recent works from the perspectives of data sources, methods and computation resources. The following are two specific examples that address online operational challenges in the grid.
\begin{table*}[htbp]
	\caption{Tri-factors of AI Solutions to Grid Monitoring, Control, and Protection}
	\label{tab:grid_online}
	\centering

	\begin{tabular}{l|l|l|l}
	\hline

	\hline
	\textbf{Task [Ref.]} & \textbf{Data Sources {(Availability)}} &\textbf{Computation Resource}& \textbf{AI Method}  \\
	\hline
	     \begin{tabular}[c]{@{}l@{}}Cyberattack detection \cite{8956077}\end{tabular}& \begin{tabular}[c]{@{}l@{}}13-, 123- IEEE {(N)}\end{tabular}  &\begin{tabular}[c]{@{}l@{}} -\end{tabular}& \begin{tabular}[c]{@{}l@{}}Bayes classifier\end{tabular}\\
	\hline
	     \begin{tabular}[c]{@{}l@{}}Actuator placement \cite{8625444
}
\end{tabular}& \begin{tabular}[c]{@{}l@{}}118-, 123- IEEE {(N)}\end{tabular} &\begin{tabular}[c]{@{}l@{}}2-core Xeon  CPU, 32GB RAM  \end{tabular}& \begin{tabular}[c]{@{}l@{}}K-means clustering\end{tabular} \\
\hline
	     \begin{tabular}[c]{@{}l@{}}Frequency prediction\\ assessment and control \cite{ 8723612}
\end{tabular}&\begin{tabular}[c]{@{}l@{}}140-NPCC {(N)}\end{tabular}& \begin{tabular}[c]{@{}l@{}}i5-5200U CPU, 8GB RAM\end{tabular} & \begin{tabular}[c]{@{}l@{}}ELM\end{tabular} \\
\hline
	     \begin{tabular}[c]{@{}l@{}}Emergency control \cite{ 8787888}
\end{tabular}& \begin{tabular}[c]{@{}l@{}}39-IEEE {(N)}\end{tabular} &\begin{tabular}[c]{@{}l@{}}AMD Opteron CPU, 64GB RAM  \end{tabular}& \begin{tabular}[c]{@{}l@{}}Deep reinforcement \\learning\end{tabular} \\

\hline
	     \begin{tabular}[c]{@{}l@{}}Under-voltage\\ load shedding \cite{9312447}
\end{tabular}& \begin{tabular}[c]{@{}l@{}}77-Nordic {(N)}\end{tabular}&\begin{tabular}[c]{@{}l@{}}8-core i7-7700 CPU; \\Nvidia GTX-1080 GPU  \end{tabular} & \begin{tabular}[c]{@{}l@{}}Deep feedback\\ learning machine\end{tabular} \\

\hline
	     \begin{tabular}[c]{@{}l@{}}Transient stability prediction \cite{ 8920121}
\end{tabular}& \begin{tabular}[c]{@{}l@{}}39-IEEE {(N)}\end{tabular} &\begin{tabular}[c]{@{}l@{}}i7-7700 CPU, 32GB RAM; \\Nvidia GTX-1080 GPU  \end{tabular}& \begin{tabular}[c]{@{}l@{}}CNN\end{tabular} \\
\hline
	     \begin{tabular}[c]{@{}l@{}}Transient stability prediction \\and control\cite{8447254}
\end{tabular}& \begin{tabular}[c]{@{}l@{}}39-IEEE {(N)}\end{tabular} &\begin{tabular}[c]{@{}l@{}}- \end{tabular}& \begin{tabular}[c]{@{}l@{}}AdaBoost \end{tabular} \\

\hline
	     \begin{tabular}[c]{@{}l@{}}Generator dynamic behavior \\prediction \cite{9193985}
\end{tabular}& \begin{tabular}[c]{@{}l@{}}39-, 68-, 140-, 145- IEEE {(N)}\end{tabular} &\begin{tabular}[c]{@{}l@{}}Intel CPU, 16GB RAM  \end{tabular}& \begin{tabular}[c]{@{}l@{}}Ensemble decision trees\end{tabular} \\

\hline
	     \begin{tabular}[c]{@{}l@{}}Voltage stability margin \\prediction \cite{9381611}
\end{tabular}& \begin{tabular}[c]{@{}l@{}}118-NREL; Taiwan Power Systems {(N)}\end{tabular} &\begin{tabular}[c]{@{}l@{}}8-core CPU, 16GB RAM \end{tabular}& \begin{tabular}[c]{@{}l@{}}ELM \end{tabular} \\

\hline
	     \begin{tabular}[c]{@{}l@{}}Training data preparation\\ for transient stability prediction \cite{9557843}
\end{tabular}& \begin{tabular}[c]{@{}l@{}}39-IEEE;  
2417-bus GD Power Grid\\ in South China 
{(N)}\end{tabular}  &\begin{tabular}[c]{@{}l@{}}8-core i7 CPU, 8GB RAM  \end{tabular}& \begin{tabular}[c]{@{}l@{}}Semi-supervised\\ ensemble learning \end{tabular}\\

\hline
	     \begin{tabular}[c]{@{}l@{}}Representative state selection\\ for security prediction \cite{8418852}
\end{tabular}& \begin{tabular}[c]{@{}l@{}}118-IEEE
{(N)}\end{tabular}  &\begin{tabular}[c]{@{}l@{}}8-core Intel Xeon PC\end{tabular}& \begin{tabular}[c]{@{}l@{}}C-Vine pair-copula\\ decomposition; PCA \end{tabular}\\

 \hline
	     \begin{tabular}[c]{@{}l@{}}Hydrostatic tidal turbine \\(HTT) control \cite{8618342}
\end{tabular}& \begin{tabular}[c]{@{}l@{}}HTT simscape model
{(N)}\end{tabular} &\begin{tabular}[c]{@{}l@{}}-\end{tabular}& \begin{tabular}[c]{@{}l@{}}Extreme learning machine
 \end{tabular} \\
 
 \hline
	     \begin{tabular}[c]{@{}l@{}}Cyber-physical anomaly detection \cite{9094728}
\end{tabular}& \begin{tabular}[c]{@{}l@{}}HiL test of Kundur two area system
{(N)}\end{tabular} &\begin{tabular}[c]{@{}l@{}}-\end{tabular}& \begin{tabular}[c]{@{}l@{}}Decision tree; KNN
 \end{tabular} \\
 
 \hline
	     \begin{tabular}[c]{@{}l@{}}Inverter control \cite{8846225}
\end{tabular}& \begin{tabular}[c]{@{}l@{}}123-IEEE
{(N)}\end{tabular} &\begin{tabular}[c]{@{}l@{}}i5 2.4 GHz, 8GB RAM \end{tabular}& \begin{tabular}[c]{@{}l@{}}Support vector machine\\(SVM)
 \end{tabular} \\
 
  \hline
	     \begin{tabular}[c]{@{}l@{}}Transient stability prediction \cite{9468376}
\end{tabular}& \begin{tabular}[c]{@{}l@{}}39-IEEE; 2417-bus GD Power Grid \\in South China
{(N)}\end{tabular} &\begin{tabular}[c]{@{}l@{}}8-core i7 CPU, 8GB RAM  \end{tabular}& \begin{tabular}[c]{@{}l@{}}Time series \\shapelet learning 
 \end{tabular} \\
 
  \hline
	     \begin{tabular}[c]{@{}l@{}}Line outage detection \cite{8747534}
\end{tabular}& \begin{tabular}[c]{@{}l@{}}30-, 118-, 300- IEEE  
{(N)}\end{tabular} &\begin{tabular}[c]{@{}l@{}}i7 CPU, 8GB RAM \end{tabular}& \begin{tabular}[c]{@{}l@{}}Learning-to-infer
 \end{tabular} \\
 
 
 \hline
	     \begin{tabular}[c]{@{}l@{}}Dynamic security prediction\cite{9105102}
\end{tabular}& \begin{tabular}[c]{@{}l@{}}39-IEEE  
{(N)}\end{tabular} &-& \begin{tabular}[c]{@{}l@{}}CNN, LSTM network
 \end{tabular} \\

\hline
	     \begin{tabular}[c]{@{}l@{}}Distribution system state estimation \cite{8693907}
\end{tabular}& \begin{tabular}[c]{@{}l@{}}37-IEEE  
{(N)}\end{tabular} &-& \begin{tabular}[c]{@{}l@{}}NN
 \end{tabular} \\
 
 \hline
	     \begin{tabular}[c]{@{}l@{}}Local control design for \\active
distribution grids \cite{8667699}

\end{tabular}& \begin{tabular}[c]{@{}l@{}}Typical European radial LV grid 
{(N)}\end{tabular} &\begin{tabular}[c]{@{}l@{}}Intel Core i7-2600 CPU \\16GB RAM \end{tabular}& \begin{tabular}[c]{@{}l@{}}Regression; SVM
 \end{tabular} \\
 
 \hline
	     \begin{tabular}[c]{@{}l@{}}Anomaly detection, \\localization and classification\cite{9335975}
\end{tabular}& \begin{tabular}[c]{@{}l@{}}14-, 39- IEEE 
{(N)}\end{tabular} &\begin{tabular}[c]{@{}l@{}}Multiple CPUs \end{tabular}& \begin{tabular}[c]{@{}l@{}}Autoencoder
 \end{tabular} \\
 
 \hline
	     \begin{tabular}[c]{@{}l@{}}Volt-VAR optimization \cite{9143169}
\end{tabular}& \begin{tabular}[c]{@{}l@{}}13-, 123- IEEE 
{(N)}\end{tabular} &\begin{tabular}[c]{@{}l@{}}i5 CPU, 8GB RAM \end{tabular}& \begin{tabular}[c]{@{}l@{}}Reinforcement learning
 \end{tabular} \\
 
 \hline
	     \begin{tabular}[c]{@{}l@{}}Cyber anomaly detection \cite{8977490}
\end{tabular}& \begin{tabular}[c]{@{}l@{}}39-IEEE 
{(N)}\end{tabular} &\begin{tabular}[c]{@{}l@{}}- \end{tabular}& \begin{tabular}[c]{@{}l@{}}SVM, Decision tree
 \end{tabular} \\
 
 \hline
	     \begin{tabular}[c]{@{}l@{}}Faulted line localization \cite{8718345}
\end{tabular}& \begin{tabular}[c]{@{}l@{}}39-, 68- IEEE 
{(N)}\end{tabular} &\begin{tabular}[c]{@{}l@{}}i7 CPU, 32GB RAM  \end{tabular}& \begin{tabular}[c]{@{}l@{}}CNN
 \end{tabular} \\

 \hline
	     \begin{tabular}[c]{@{}l@{}}Feature extraction for\\ Security assessment \cite{8477148}
\end{tabular}& \begin{tabular}[c]{@{}l@{}}118-IEEE 
{(N)}\end{tabular} &\begin{tabular}[c]{@{}l@{}}8-core Intel Xeon PC \end{tabular}& \begin{tabular}[c]{@{}l@{}}Deep learning
 \end{tabular} \\
\hline
	     \begin{tabular}[c]{@{}l@{}}Building Energy Optimization \cite{8356086}
\end{tabular}& \begin{tabular}[c]{@{}l@{}}Real-world data from Pecan Street Inc.
{(N)}\end{tabular} &\begin{tabular}[c]{@{}l@{}}-\end{tabular}& \begin{tabular}[c]{@{}l@{}}Reinforcement learning
 \end{tabular} \\
 \hline
	     \begin{tabular}[c]{@{}l@{}}PV frequency control \cite{9296965 }
\end{tabular}& \begin{tabular}[c]{@{}l@{}}6,102-bus system;
HiL tests
{(N)}\end{tabular} &\begin{tabular}[c]{@{}l@{}}-\end{tabular}& \begin{tabular}[c]{@{}l@{}}Multivariate random\\forest regression 
 \end{tabular} \\

 \hline
	     \begin{tabular}[c]{@{}l@{}}Wind turbine control \cite{9470951}
\end{tabular}& \begin{tabular}[c]{@{}l@{}}
HiL tests
{(N)}\end{tabular} &\begin{tabular}[c]{@{}l@{}}-\end{tabular}& \begin{tabular}[c]{@{}l@{}}Cascade-forward\\neural network 
 \end{tabular} \\
 
 \hline
	     \begin{tabular}[c]{@{}l@{}}Non-intrusive load monitoring \cite{ 8437176 }
\end{tabular}& \begin{tabular}[c]{@{}l@{}}
Reference Energy Disaggregation Dataset
{(Y)}\end{tabular} &\begin{tabular}[c]{@{}l@{}}i7-4770 CPU, 12GB RAM\end{tabular}& \begin{tabular}[c]{@{}l@{}}Semi-supervised multi-\\label algorithms
 \end{tabular} \\
 
 \hline
	     \begin{tabular}[c]{@{}l@{}}Building occupancy detection \cite{9043691}
\end{tabular}& \begin{tabular}[c]{@{}l@{}}
Electricity consumption and occupancy \\(ECO) dataset; Building-level fully labeled\\ Electricity disaggregation dataset;
UMass \\Smart Home Dataset 
Almanac of Minutely \\Power Dataset

{(Y)}\end{tabular} &\begin{tabular}[c]{@{}l@{}}Intel Xeon E5-2603 CPU; \\Nvidia TITAN V GPU\end{tabular}& \begin{tabular}[c]{@{}l@{}}CNN, LSTM network
 \end{tabular} \\
 
 \hline
	     \begin{tabular}[c]{@{}l@{}}Outage duration prediction \cite{8421644}
\end{tabular}& \begin{tabular}[c]{@{}l@{}}
E15-year outage records
{(N)}\end{tabular} &\begin{tabular}[c]{@{}l@{}}-\end{tabular}& \begin{tabular}[c]{@{}l@{}}Natural language processing 
 \end{tabular} \\
  \hline
	     \begin{tabular}[c]{@{}l@{}}Event detection and classification \cite{9619930}
\end{tabular}& \begin{tabular}[c]{@{}l@{}}
HiL tests
{(N)}\end{tabular} &\begin{tabular}[c]{@{}l@{}}i7-9700K CPU;\\Nvidia GTX 2080Ti\\ GPU\end{tabular}& \begin{tabular}[c]{@{}l@{}}CNN
 \end{tabular} \\
 
 \hline
	     \begin{tabular}[c]{@{}l@{}}Cyber attack detection \cite{ 9559412 }
\end{tabular}& \begin{tabular}[c]{@{}l@{}}
57-, 118-, 300- IEEE
{(N)}\end{tabular} &\begin{tabular}[c]{@{}l@{}}i9-8950 HK CPU 2.90GHz \\Nvidia GeForce RTX \\2070 GPU \end{tabular}& \begin{tabular}[c]{@{}l@{}}Graphic Neural Network
 \end{tabular} \\
 
 \hline
	     \begin{tabular}[c]{@{}l@{}}Distribution system topology\\ identification \cite{9573436}
\end{tabular}& \begin{tabular}[c]{@{}l@{}}
33-IEEE; 135-bus, \\874-bus systems
{(N)}\end{tabular} &\begin{tabular}[c]{@{}l@{}}-\end{tabular}& \begin{tabular}[c]{@{}l@{}}Split expectation maximization
 \end{tabular} \\
 
 \hline
	     \begin{tabular}[c]{@{}l@{}}Grid restoration \cite{9451602 }
\end{tabular}& \begin{tabular}[c]{@{}l@{}}
70-bus 4-feeder system
{(N)}\end{tabular} &\begin{tabular}[c]{@{}l@{}}i5 CPU, 8GB RAM \end{tabular}& \begin{tabular}[c]{@{}l@{}}Regression
 \end{tabular} \\
 
	\hline

	\hline
	\end{tabular}
\end{table*}
\subsubsection{Forced oscillation localization based on robust principal component analysis (RPCA)}
\begin{figure}[thb]
    \centering
    \includegraphics[width = 3.5in]{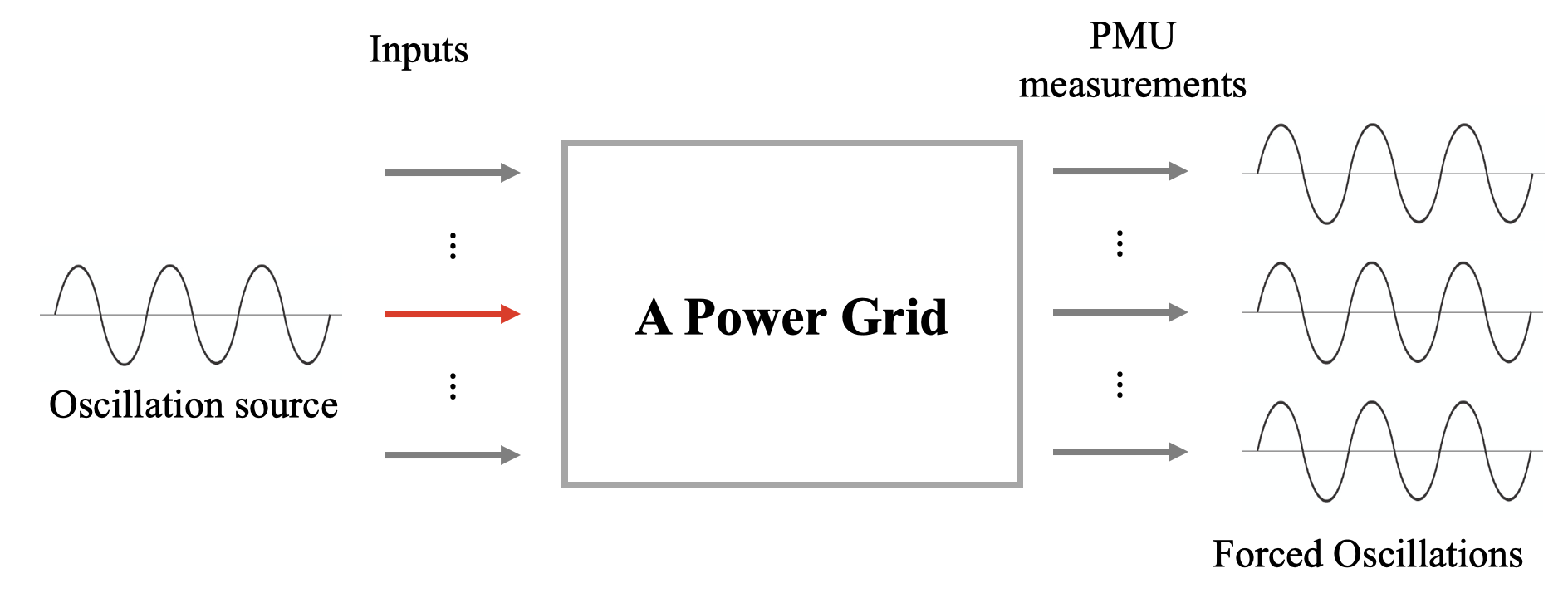}
    \caption{Forced oscillation mechanism}
    \label{fig:FO}
\end{figure}

Forced oscillations are one type of the critical phenomena that concern system operators, because these oscillations may cause large-scale blackouts and decrease the lifespans of power grid components\cite{9043670}. Fig. \ref{fig:FO} illustrates the mechanism of forced oscillations. Let us consider a power grid as a blackbox with some inputs and outputs, as shown in Fig. \ref{fig:FO}. The inputs can be thought of as setpoints of generators, while the outputs are PMU measurements. If one of the inputs varies periodically, oscillations can be observed in the PMU measurements. These oscillations are termed ``the forced oscillations," and the periodic input is called the source of the forced oscillations. Different PMU measurements have different geographical distances from the oscillation source. The objective of the forced oscillation localization is to pinpoint which PMU measurements are close to the oscillation source, based only on the PMU data without information on the inputs and the power grid models.

\begin{figure}[thb]
	\centering
	\includegraphics[width = 2.5in]{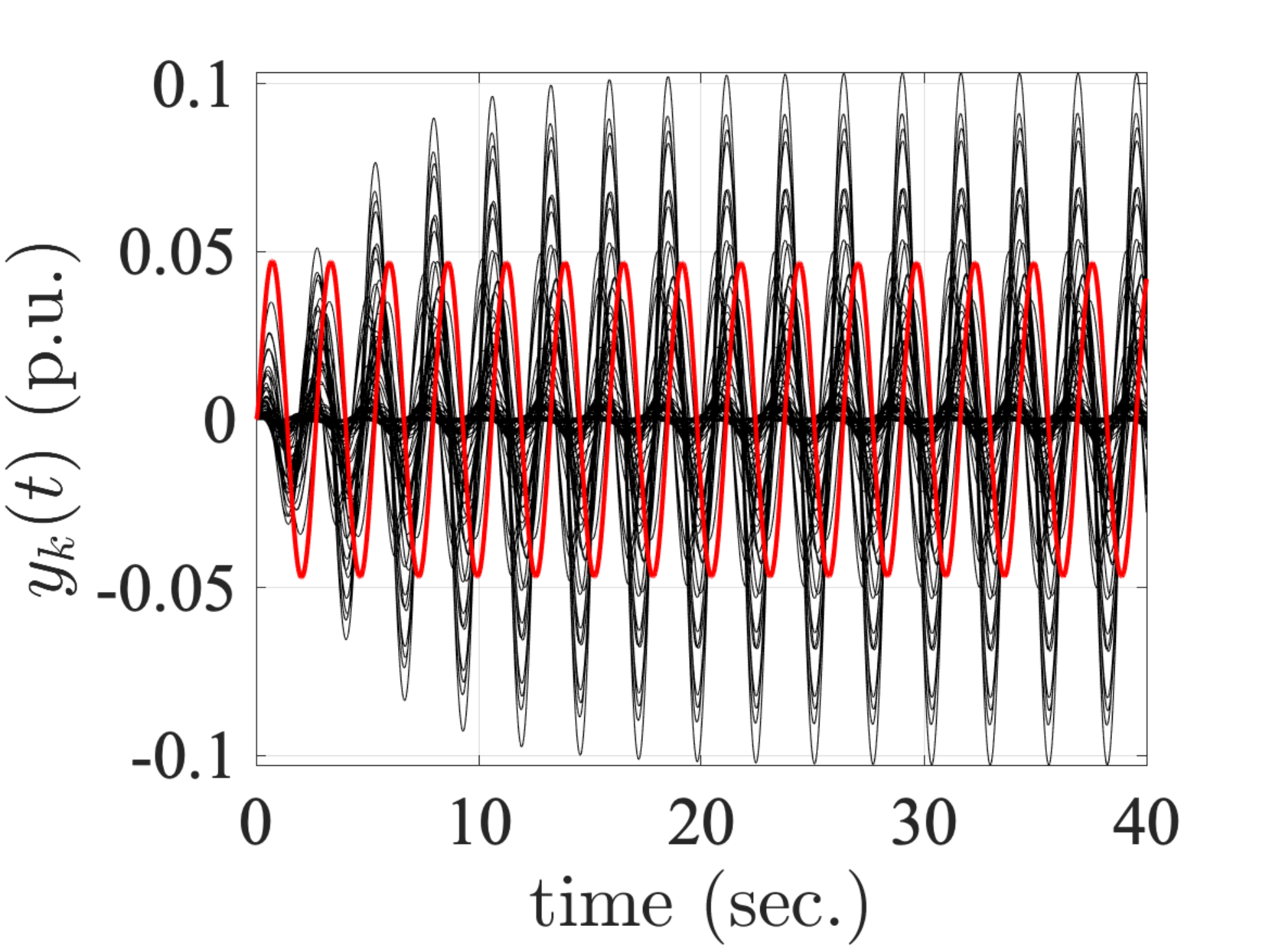}
	\caption{Voltage deviations in a counter-intuitive case: the red curve is the voltage deviation closest to the oscillation source; the black curves are other voltage deviation measurements. (Source: Fig. 1 of reference \cite{9043670} \copyright IEEE 2021)}
	\label{fig:counter-intuitive}
\end{figure}

\begin{figure*}[thb] 
				\centering
				\subfloat[]{\includegraphics[width=0.32\textwidth]{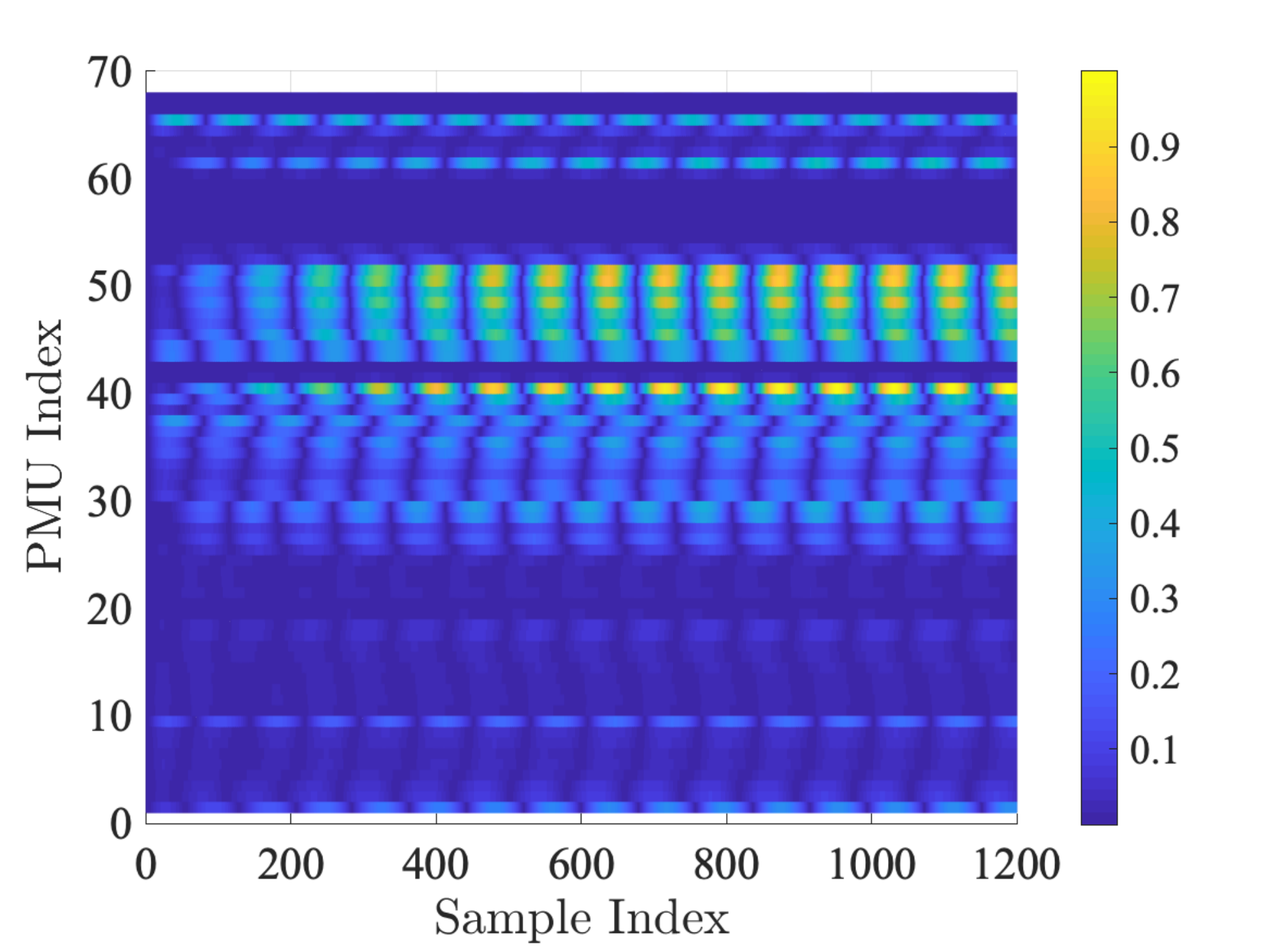}}
				\hfil
				\subfloat[]{\includegraphics[width=0.32\textwidth]{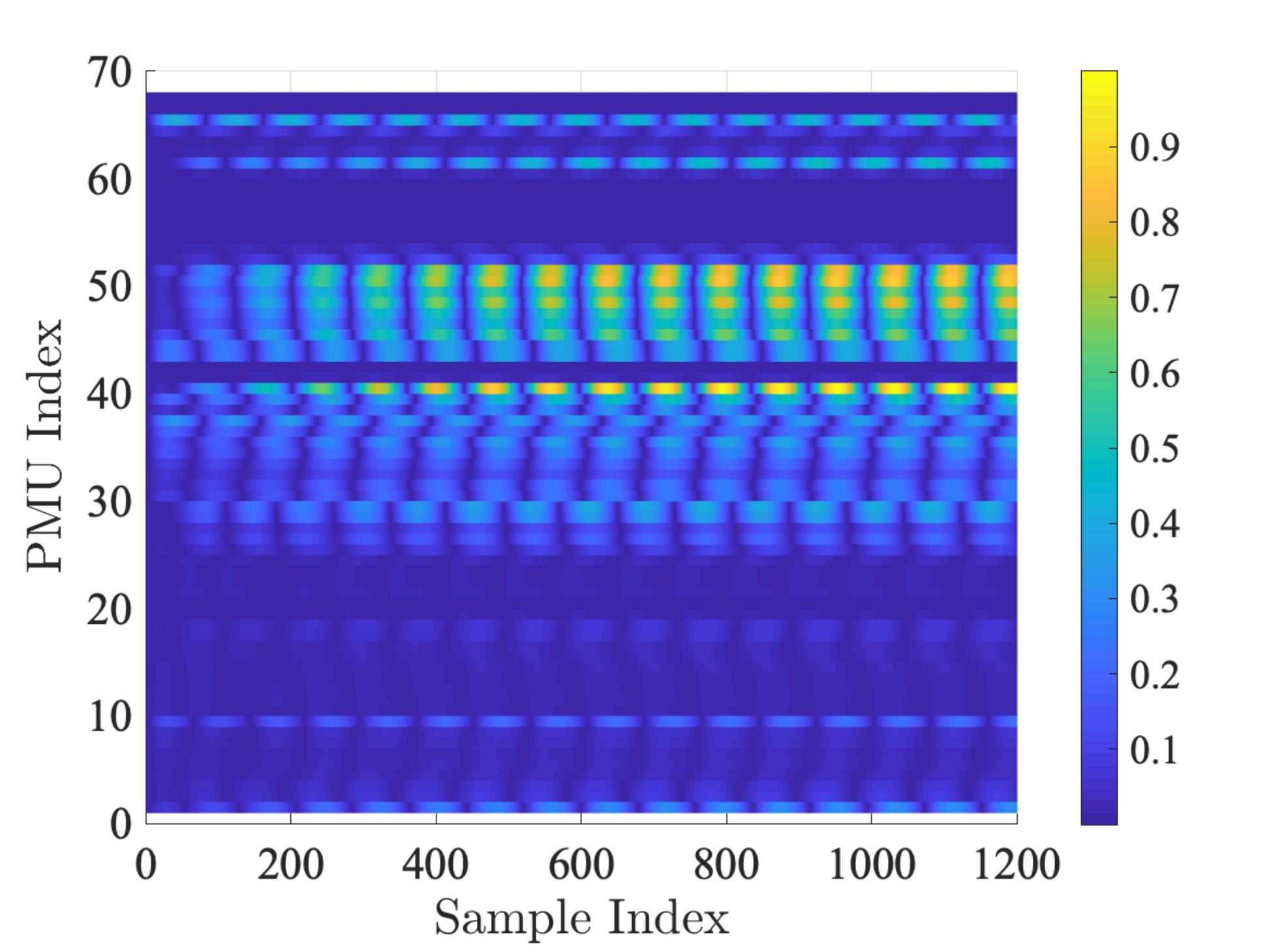}}
				\hfil
				\subfloat[]{\includegraphics[width=0.32\textwidth]{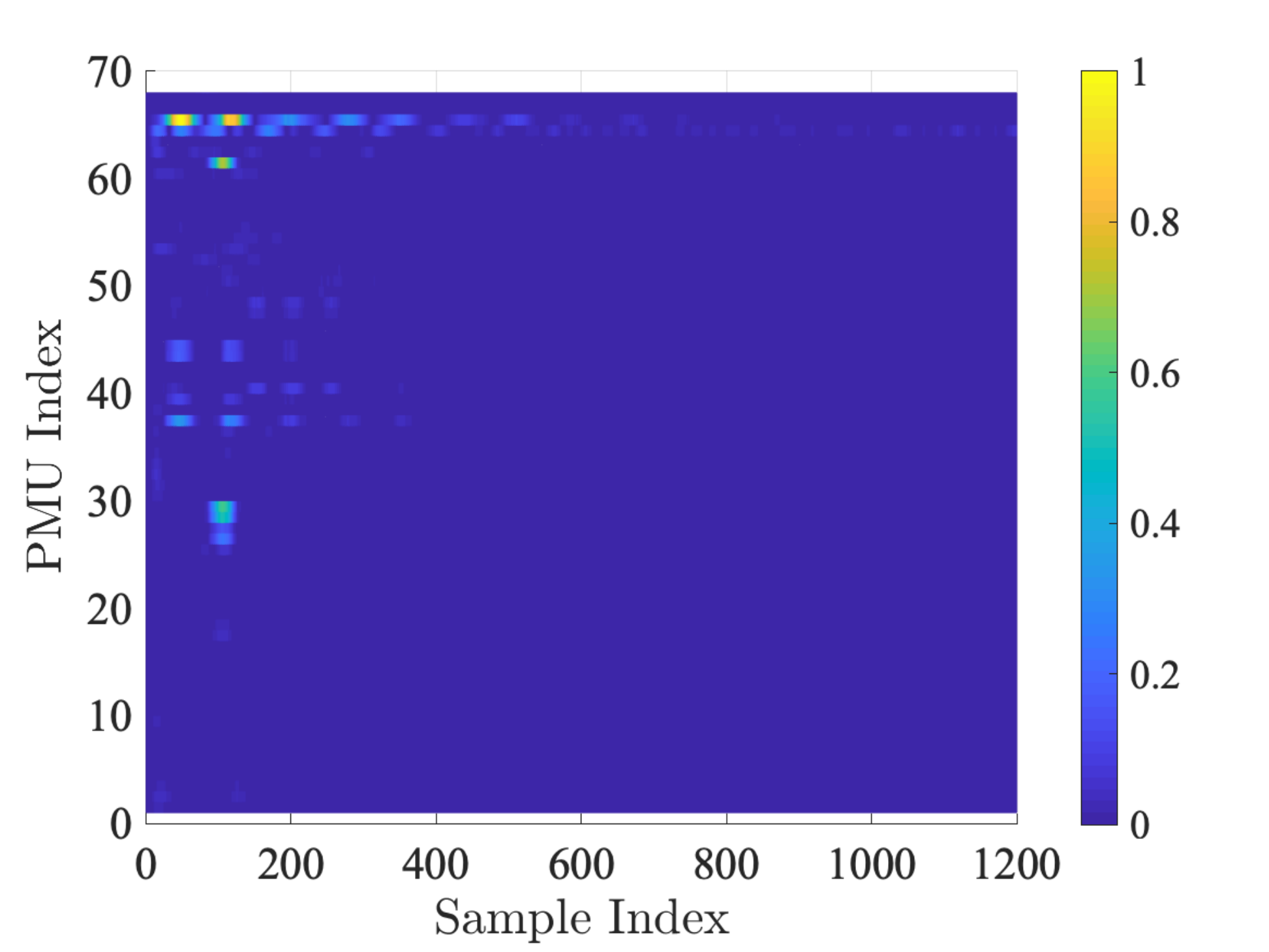}}
				\hfil
				\caption{Illustration of the RPCA-based source localization algorithm:  (a) the measurement matrix; (b) the low-rank matrix; and (c) the sparse matrix. The (normalized) magnitudes of matrix entries are color-coded.The measurement closest to the source can be tracked by identifying the largest absolute entry in the sparse matrix, i.e., the entry with the brightest color. (Source: Fig. 2 of reference \cite{9043670} \copyright IEEE 2021)}. \label{fig:RPCA_visualization} 
			\end{figure*}

Locating the oscillation source is a challenging task, because the measurement closest to the source may not exhibit the largest oscillations. Fig. \ref{fig:counter-intuitive} shows such a counter-intuitive case in which the measurement (the red curve) closest to the oscillation source does not exhibit the largest oscillation magnitude. Reference \cite{7401141} reports a real-world, counter-intuitive case in which the distance between the source and the measurement exhibiting large oscillations is more than 1100 miles \cite{9043670}. In reference \cite{9043670}, Huang \emph{et al.} formulate the forced oscillation localization as decomposing the measurement matrix $Y_t$ into a low-rank matrix $L_t$ and a sparse matrix $S_t$, namely $Y_t=L_t+S_t$. This matrix decomposition problem can be solved by RPCA as shown in \eqref{eq:RPCA},

\begin{equation}\label{eq:RPCA}
    \min_{S_t}\quad \|Y_t-S_t\|_{*}+\epsilon\|S_t\|_1,
\end{equation}
where $\|\cdot\|_*$ and $\|\cdot\|_1$ denote the nuclear norm and $l_1$ norm, respectively, $Y_t$ represents a measurement matrix up to time $t$ where each row of the matrix represents a time series from one PMU, and $S_t$ is the corresponding approximate sparse matrix.
Fig.~\ref{fig:RPCA_visualization} visualizes matrices $Y_t$, $L_t$, and $S_t$, respectively. The computation complexity analysis of RPCA is reported in \cite{candes2011robust}. The measurement near the source can be located by identifying the largest absolute element in the sparse matrix. Reference \cite{9043670} also provides a possible interpretation to justify the effectiveness the RPCA-based source localization algorithm. In reference \cite{9043670}, the authors create $44$ counter-intuitive cases in an open-source, benchmark system. The RPCA-based algorithm can pinpoint the sources in $43$ cases, and in the wrong case, the algorithm can narrow the searching scope \cite{9043670}. However, when the RPCA can exactly locate the true source remains an open-ended question.

\subsubsection{Reinforcement learning (RL)-based protection scheme for renewable-rich distribution systems}
\begin{figure}[htb]
    \centering
    \includegraphics[width = 3.5in]{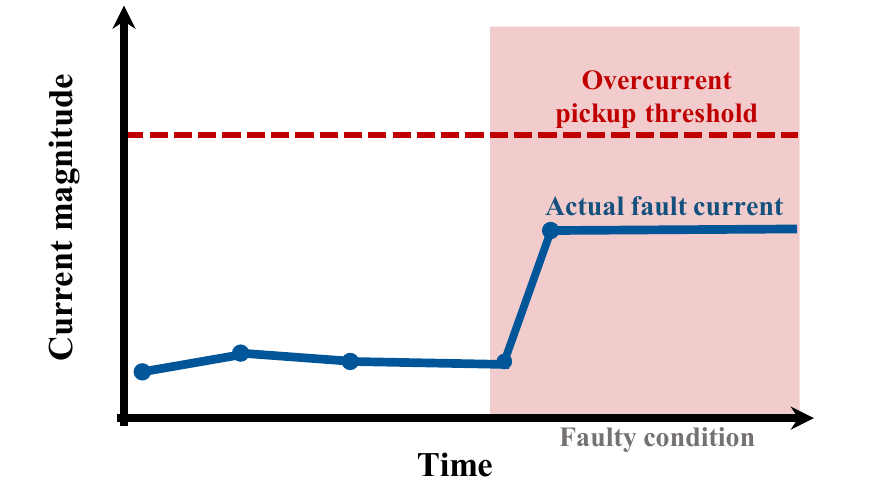}
    \caption{The conventional threshold-based protection scheme may fail due to low fault current. (Modified from source: Fig. 1 of \cite{wu2020deep})}
    \label{fig:protection}
\end{figure}

The conventional protection paradigm in distribution systems has been challenged by the increasing amount of DERs. Fig. \ref{fig:protection} presents the overcurrent protection scheme that is widely deployed in power distribution systems. Such a protection scheme will trip the line once the line current exceeds a threshold value, e.g., $5$ times the current $I_0$ under normal conditions. However, if a DER is installed nearby, it may decrease the fault current by injecting reverse power flow. As a consequence, the current under the faulty condition might be much less than the relay threshold. 

\begin{figure}[htb]
    \centering
    \includegraphics[width = 3.5in]{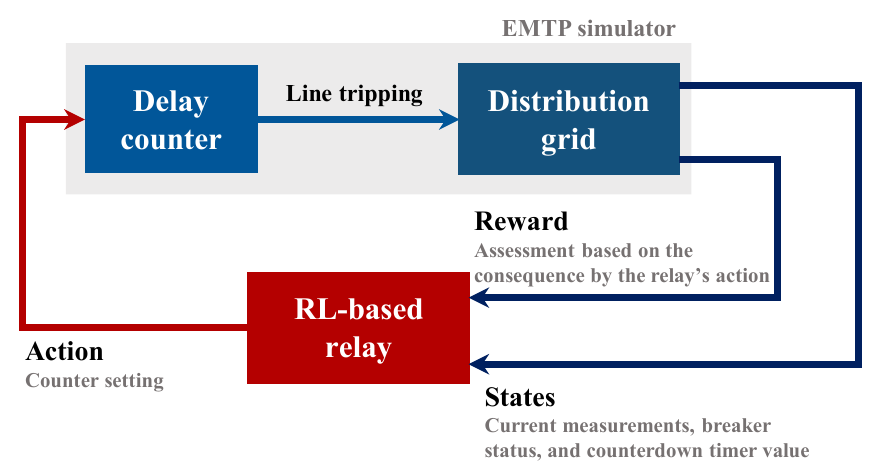}
    \caption{Obtaining the $Q$ function for the RL-based relay: the optimal policy embedded in the RL-based relay is obtained by interacting with a distribution system simulator.}
    \label{fig:RL_training}
\end{figure}

In order to address the protection challenges in a renewable-rich distribution system, reference \cite{wu2020deep} places the protection problem into a RL framework (Fig. \ref{fig:RL_training}) in which the protection scheme is learned by interacting with a distribution system simulator. In the RL framework, the distribution system is modeled by a Markov decision process (MDP) described by states $s\in\mathcal{S}$, actions $a\in\mathcal{A}$, a reward function $r(s, a)$, transition probability $P$, and a user-defined discount factor $\beta\in(0,1]$. The implication of the states, action, and reward function in the protection problem are annotated in Figure \ref{fig:protection}. In particular, the state $s_{i,t}$ and action $a_{i,t}$ of relay $i$ at time $t$ are defined by
\begin{subequations}\label{eq:RL_s_a}
\begin{align}
    &s_{i,t}=\{s_{i,t}^\text{c},s_{i,t}^\text{b},s_{i,t}^\text{d}\},\\
    &a_{i,t}=\{a_{i,t}^\text{set},a_{i,t}^\text{d},a_{i,t}^\text{reset}\},
\end{align}
\end{subequations}
where $s_{i,t}^\text{c}$ represents local current measurements, $s_{i,t}^\text{b}$ represents the status of the local breaker, $s_{i,t}^\text{c}$ represents the value of the countdown timer, $a_{i,t}^\text{set}$ represents the action of triggering the countdown timer, $a_{i,t}^\text{d}$ represents the action of decreasing the value of the counter by one, and $a_{i,t}^\text{reset}$ represents the action of resetting the counter. The reward function gives deterministic positive rewards to the tripping action under fault conditions and stay-in-silence action under normal condition, and it gives negative rewards to malfunctions.
The transition probability is determined by the distribution system; in practice it is unknown. The optimal action $a^*(s)$ at state $s$ is obtained by
\begin{subequations}\label{eq:BellmanEq}
\begin{align}
    &Q(s, a) = \mathbb{E}\left(r(s,a) + \beta\max_{a'\in \mathcal{A}}Q(s', a')\right),\\
    &a^*(s) = \arg \max_{a'\in\mathcal{A}}Q(s, a'),
\end{align}
\end{subequations}
where $\mathbb{E}(\cdot)$ is the expectation operator; $a'$ is the possible next-step action; and $s'$ is the next-step state given the current state and action; it is determined by the distribution system. In \cite{wu2020deep}, the $Q$ function in \eqref{eq:BellmanEq} is approximated by an NN. The NN's parameters are learned by a sequence of $\{s, a, r, s'\}$ observations from the framework shown in Fig. \ref{fig:RL_training}. The dataset reported in \cite{wu2021pyprod} can be used for training the algorithm. The simulation results in \cite{wu2020deep} suggests that the failure rate of the RL-based relay is only $0.32\%$ in a distribution system with $30\%$ DER penetration, whereas the conventional overcurrent relay has a much higher failure rate, i.e., $15.46\%$, under the same condition. One future direction of this work is to investigate a rigorous convergence guarantee for the sequential reinforcement learning algorithm \cite{wu2020deep}.

To summarize this section, we provide two-fold guidance on applying use-inspired AI methods in power systems.
First, it is critical to find appropriate application scenarios that take precedence over proposing innovative methodology. With deep neural networks as representatives, current AI techniques that are essentially model-agnostic function approximators usually present outperforming performance in application scenarios where there is only heuristic experience with no clear first-principle physical model, such as in load and renewable prediction. The illustrated neural network-based Lyapunov function~\cite{9559389} is another example. Although a Lyapunov function itself has rigorous definition, there is no traditional cost-effective analytical or numerical way to construct such a function for a large-scale real-world dynamical system, in which neural networks can provide an alternative effective solution.
Second, it is desirable to intelligently and insightfully formulate critical challenges in traditional power systems into AI-friendly formats. Consider illustrated forced oscillation source localization~\cite{9043670} as one example. Intuitively, it can be formulated as a typical classification problem by taking system global states as inputs and discrete location labels as outputs. However, formulated as a matrix decomposition problem, this problem can be solved by RPCA that is commonly used for image processing, which has both outperforming accuracy and explainability.

\section{Use Cases of Industry Adoption}\label{sec:use cases}
As more measurement data and data-driven algorithms become available, the power industry continues to adapt and improve operations by leveraging new technology and systems that enable it to meet and exceed customer expectations. 
This section presents some industry use cases to illustrate the continuing adoption of machine learning techniques by
Oncor, a regulated utility that operates operates the largest distribution and transmission system in Texas. The following use cases were selected to show instances of AI adoption with relatively high maturity. In addition, we illustrate use cases (e.g., asset management) that are not considered in power-systems research, but that are essential for business operations with physical devices spread over large distances.
All use case development is based on business needs and the value of the investment must be justified before a use case is developed, even if data are readily available.
Moreover, the value-add of some high performance algorithms in many cases may not offset the maintenance cost required to keep such models operating properly (e.g., due to model drift).
Table~\ref{tab:usecases} provides a brief introduction to the industry use cases that will be described in detail.
Because some use cases involve proprietary information, details about pre- and post-processing steps and model accuracy level will not be disclosed.

In many industry use-cases, the methods currently used may appear simplistic compared to the latest research; however, these use-cases are of high value, and large amounts of data are readily available. 
Utilities usually have multiple databases for various systems, such as outage management, advanced metering, work orders, geographical and meteorological data, and financial info. 
An essential challenge for conducting any big data analysis is to unify this data and enforce consistent formats for each data type.
At Oncor, a datalake was created to consolidate the data needed for analytics. The datalake replicates data from all of Oncor's operational databases. In addition to supporting uniformity, this approach also minimizes stress on operational databases because they are accessed only during each scheduled copy rather than whenever an analyst makes a query.

As the industry continues to adopt machine learning continues, and available platforms become more mature, advanced techniques will be more feasible at lower cost; these will be necessary to address more complex problems in power systems.
Most importantly, collaboration between practitioners and researchers must intensify to achieve efficient and continuous adoption.

\begin{table*}
\caption{Industry data-driven use cases}
\label{tab:usecases}

\begin{adjustbox}{width=\textwidth}
\begin{tabular}{llccccccccc}  \hline
\multirow{2}{*}{Application}  & \multicolumn{3}{c}{Algorithm} & \multicolumn{4}{c}{Data} &\multirow{2}{*}{Computing Platform} \\\cmidrule(lr){2-4}
\cmidrule(lr){5-8} \cmidrule(lr){10-11}
& Main method & Type & Accuracy & Source & Resolution & \# Sample & Dimension &  & \\\hline
 Transformer voltage anomaly classification & Change point detection  & Offline & 94\% & AMI & 15min & $3.7*10^6$ & $672\times1$ & Spark*  \\ 
 Asset defect detection  & YOLOv3 & Offline & $\geq87\%$ & Aerial images &  
  - & 500 & $608\times608$ & Tesla V100 GPU \\ 
 Short-/Mid-term load forecast  & Regression trees & Offline & - & AMI & 15min & $3.7*10^6$ & $4.1*10^4\times1$ & Spark*  \\
 Cold load characterization & Linear regression & Offline & - & SCADA & 1min & 1300 & Various & - \\
 RIC Categorization & Cluster analysis & Offline & - & AMI & 15min & $4.9\times10^5$ & $8,064\times1$ &  Spark* \\
\hline
\end{tabular}
\end{adjustbox}
\\[0.25\baselineskip]
\centering{*Spark: 2 namenodes, dual Xeon-4208 (8-core); 8 datanodes, dual Xeon-5218 (16-core); 768 GB RAM per node.}
\end{table*}

\subsection{Asset Health}
For all utility companies, monitoring and maintaining their assets is critical to realizing system reliability and providing the highest quality service to their customers.
Some assets, such as distribution class transformers, can be monitored by utilizing AMI meter data, such as voltage and kWh readings.
For assets where digital measurements are not available, health monitoring may be possible by analyzing asset images using advanced image processing techniques.
Several Oncor use cases are presented below to illustrate how asset health can be monitored by utilizing machine learning methods.

As the largest utility company in the state of Texas, Oncor provides power to nearly 4~million customers through more than 1~million distribution class transformers, which can fail from damaged coils or overload degradation. 
Reactive replacement of a failed transformer can take more than 4~hours, but proactive replacements often take less than 1~hour. 
Thus, detecting failure precursors can significantly reduce both labor cost and outage time. 

\begin{figure}[tb]
\centering
\includegraphics[width=0.48\textwidth, trim=5 0 0 0, clip=true]{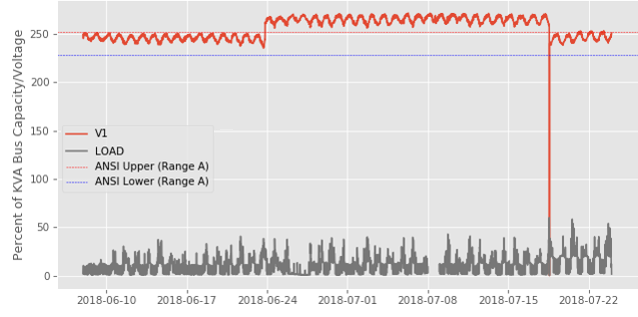}
\caption{The voltage profile of a transformer with a coil damage and subsequent replacement. Voltage profile is in red and load profile is in grey. }
\label{fig:XFMRvoltage}
\end{figure}

Fig.~\ref{fig:XFMRvoltage} shows a plot of the voltage and load measurements from a single phase 240V AMI meter. 
Both voltage ``V1" (in Volts) and load ``LOAD" (in kWh) time series, in red and grey respectively, have a 15-minute resolution.
The two horizontal lines are the upper and lower limits of the operating voltage ratings defined by the American National Standards Institute (ANSI C84.1-2020), which are $\pm5\%$ of the nominal voltage. On June 24$^\text{th}$, 2018 the voltage suddenly rose above the upper limit due to a damaged coil on the primary side of the transformer. 
The sudden drop in voltage on July 18$^\text{th}$, 2018 denotes the time of the replacement.
Typically, a transformer will not fail immediately after a coil is damaged.
Therefore, proactive replacement is realistic and valuable if a change in voltage can be detected soon enough.

After examining the pre-outage voltage profiles of all transformers replaced in Oncor's system during an 18-month period, a change point detection algorithm was designed to detect over/under voltage issues. 
A change in mean and/or variance of a meter's voltage was detected by a PySpark implementation of the functions provided in \cite{changepoint1}. 
Several post-processing steps were implemented to remove change points due to outages or temporary voltage changes. 
The thresholds for these steps were selected from the ground truth data. 
Based on the number of issues seen on the same feeder, the detected issues were then categorized into various types, such as meter, transformer, or regulation issues, to enhance the troubleshooting process of the distribution operations organization. 
The algorithm and thresholds were tuned and improved using feedback received from the field. 
Currently the voltage monitoring process runs every weekday on data from 3.7 million AMI meters. 
The weekly-average accuracy for June-November 2021 is $94\%$. 

Oncor began to monitor distribution transformer health in 2016. 
As of November 2021, 3834~issues have been resolved proactively using transformer health analysis. 
These issues include damaged transformers or meters, as well as installation, regulation, and secondary issues that affect voltage measurements. 
Proactive transformer maintenance has saved Oncor approximately $\$3.25$~million in equipment, labor and expenses as well as 5.5~million customer interruption minutes.

Another asset health use case is defective insulator detection, due to, for example lightning strikes, forceful impacts or aging. Defective insulators are hazardous to the operation of power lines and pose a risk to system reliability. Oncor has more than $18,000$~circuit-miles of transmission lines with over $500,000$ transmission insulators.
Rapid identification of damaged insulators, especially after a storm, is therefore a critical task in asset management. 
Due to the scale of Oncor's transmission system, manual inspection is infeasible.
An automated inspection method was developed that use aerial/drone images of transmission lines and convolutional neural networks.

The insulator defect detection method employs YOLOv3 (You Only Look Once, Version 3~\cite{redmon2018yolov3}), which is a real time object detection model that uses Darknet-53~\cite{darknet13} as the backbone feature extractor in a deep convolutional neural network. 
The model was initialized with YOLO's pre-trained weights using the Microsoft COCO (Common Objects in Context) dataset~\cite{lin2015microsoft} and insulator images, provided by Electric Power Research Institute (EPRI), were used for transfer learning and validation (confidential data).  

The defect detector successfully recognized the insulators in an image, pinpointed those issues of each damaged insulator, and classified the issues as either ``broken" or ``flashed." 
For the 50 testing images, each containing multiple flashed/broken locations, $100\%$ of the broken points were detected correctly and $90\%$ of the flashed points were detected. 
There were no misclassified issues.

The recent Texas House Bill 4150, also known as the ``William Thomas Heath Power Line Safety Act," which was passed through the Legislature in May 2019, requires all utilities to make regular inspections of their power lines to ensure that they comply with state and federal safety regulations. Although Oncor completes routine inspections of all transmission power lines, detailed manual inspections of all structures are time consuming, impactful to land owners and costly. In an effort to reduce resources such as on right-of-way truck traffic, another deep model was trained and applied to aerial images of the power lines. This model is being developed in stages to ultimately identify reliability risks due to structures damaged by impacts or aging.  
The first stage of this model requires Oncor to verify all structure asset information in the Oncor Transmission Information System. 
Because many transmission lines are 40+ years old, information in historical records may be inaccurate for structures where components were replaced or added after the initial installation.
Additional stages include identifying attributes that can indicate structural issues that may cause outages and affect reliability performance. 
These attributes include
\begin{itemize}
    \item Composition: wood, steel, concrete
    \item Design: H-frame, A-frame, lattice tower, multi-pole, single-pole
    \item Cross arm: beam, double-plank 
    \item Brace: V, X, knee
\end{itemize}

\begin{figure}[tb]
\centering
\includegraphics[width=0.48\textwidth]{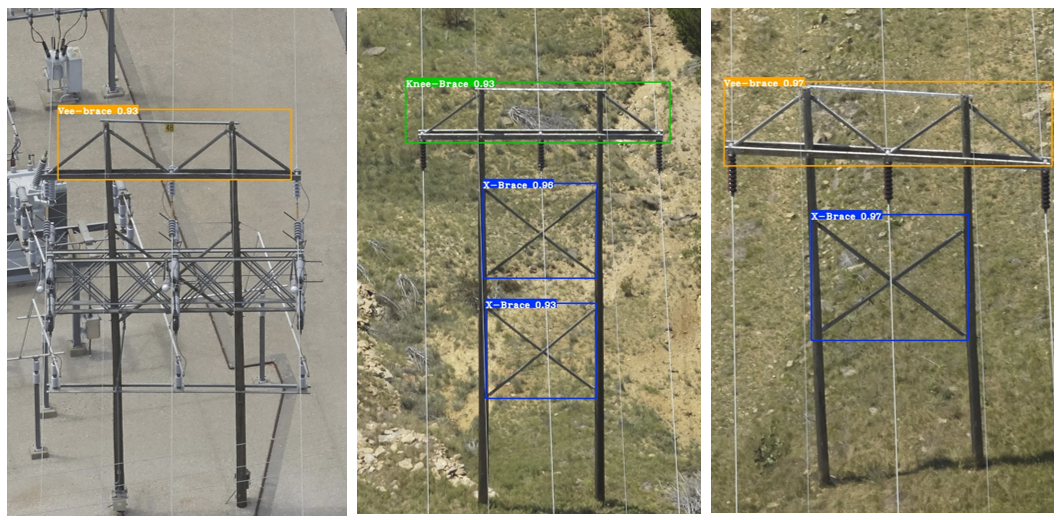}
\caption{Examples for brace type classification. Yellow boxes: V brace, green box: knee brace, blue boxes: X brace.}
\label{fig:brace}
\end{figure}

\begin{figure}[tb]
\centering
\includegraphics[width=0.48\textwidth]{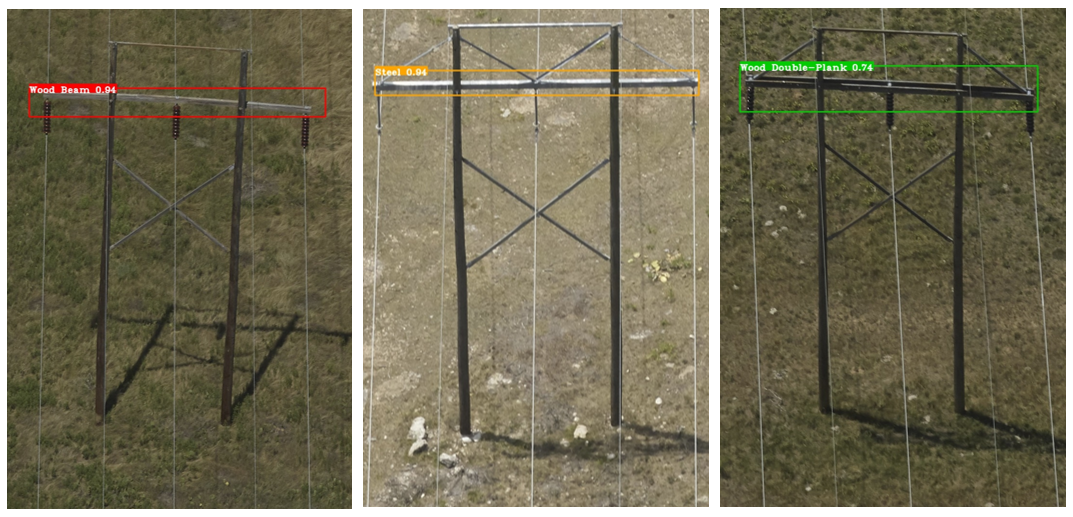}
\caption{Examples for beam type classification. Red box: wood beam, yellow box: steel beam, green box: wood double-plank.}
\label{fig:beam}
\end{figure}

The effort to classify transmission line attributes made use of YOLOv3; the initial results were promising, with accuracy rates of~$89\%$ for braces and $87\%$ for cross arms. 
Fig.~\ref{fig:brace} and Fig.~\ref{fig:beam} show several examples of successful classification results.  
As more images are labeled to augment training data, the model's performance is expected to improve; furthermore, by including images with defective structures, the system can be used to inventory components as well as their degradation levels. 

\subsection{Load Forecasting}

Load forecasting is an essential building block in operating and planning tasks in both the power industry~\cite{hong2016probabilistic} and commercial building energy~\cite{sun2020data}. 
It is needed in many decision making processes for electric energy generation, DERs management, transmission, distribution, markets, and demand-response.
The pursuit of models that can achieve accurate load forecasts for short-, mid-, and/or long-term purposes is a long standing research area with a large body of literature~\cite{gross1987short, park1991electric}.   

For utility companies, short- and mid-term load forecasts are used to plan switching operations in control centers. 
Moreover, load forecasts contribute to network reconfiguration and infrastructure development/improvement decisions.
For example, to better prepare for high power demand seasons, Oncor conducts load analyses to forecast summer and winter feeder load peaks.  
In some cases, a contingency plan will be made ahead of these peak seasons for feeders that are at risk of overload based on historical load data leveraged by analytics.  

These efforts have significantly improved Oncor’s reliability performance; there has not been a feeder lockout event due to overload since 2018.
Switching operations, however, are a major challenge for feeder load forecasting because a feeder's load can change significantly due to a load switching event (e.g. feeder reconfiguration due to an outage or planned maintenance). 
A robust model is needed to respond to these events quickly and adjust the forecasts correspondingly.
Oncor currently is developing deep learning methods to surpass the performance of the current approach.

Besides feeder load forecasts, load forecasting at any device is needed for making operational decisions in the control rooms. 
One approach is to forecast the load at each distribution transformer using AMI meter data and then aggregate at each device as needed. 
With a large quantity of distribution transformers (e.g., more that 1 million in Oncor's system), if computational power is limited, cluster analysis can be used to group transformers with similar load behaviors.
Normalization (re-scaling each load profile to range $[0,1]$) is needed before clustering so that the clustering results are affected mainly by the shape of the load profiles.
After the transformers have been assigned into clusters, load forecasts for each cluster center (the representative of all transformers in that cluster) can be obtained; they are then scaled back to each transformer's load level by undoing the normalization steps.
If distributed computing platforms are available, transformer load forecasting can be conducted by directly training individual models for every transformer, which will introduce fewer errors.

Oncor implemented a regression tree model \cite{breiman2017classification} on Spark that serves both short- and mid-term needs.
The load of a transformer is affected by both numerical and categorical factors. 
The most important numerical factors include temperature, wind speed, humidity, and solar radiation, whereas categorical factors include time of day, day of week, month, \emph{etc}.
To avoid over-fitting, the maximum numbers of layers and leaves were tuned based on model performance.

\begin{figure}[tb]
\centering
\includegraphics[width=0.48\textwidth]{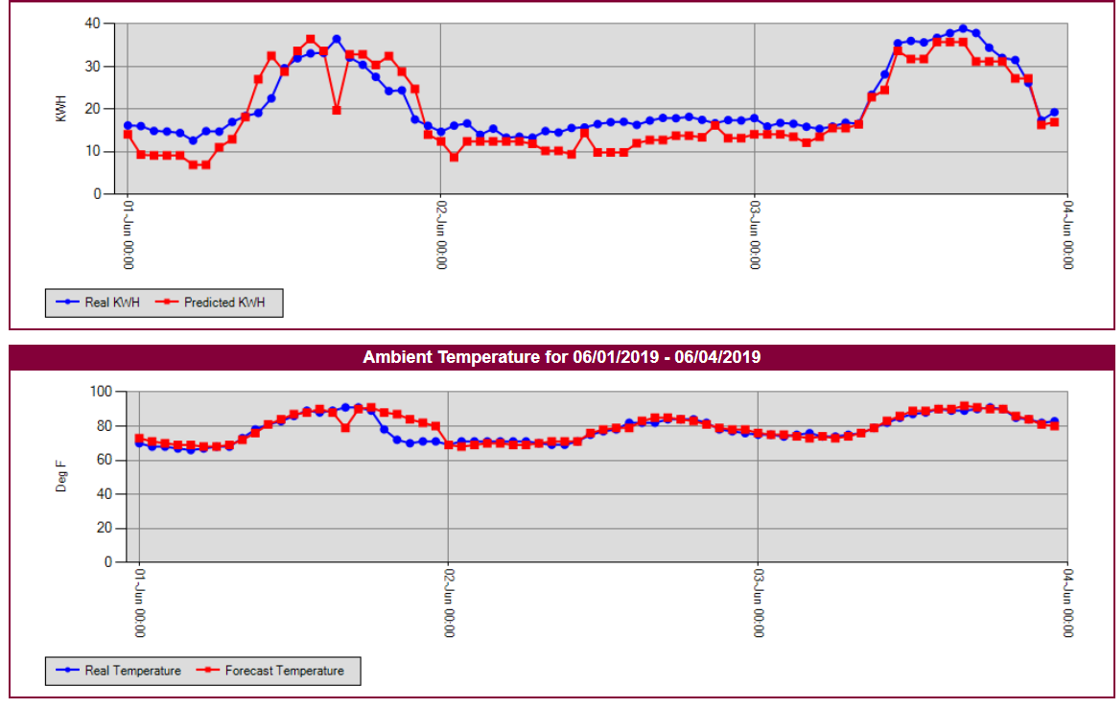}
\caption{An example of transformer load forecast results. Blue dots: actual measurements; red dots: predicted values. Top figure: predicted and actual loads; bottom figure: predicted and actual temperatures.}
\label{fig:loadforecast}
\end{figure}

Fig.~\ref{fig:loadforecast} shows an example of the hourly load forecasting results for one distribution transformer over the course of 3 days. 
The blue and red curves on the top plot give the actual and predicted load based on the predicted temperatures in the bottom plot (blue curve) using a regression tree model trained for a particular transformer.
There is a trade-off between model performance (error level) and computing time, which can be calibrated to suit shifting business needs at any given time.
This approach is able to capture non-periodic activity that sometimes deviates from the temperature as seen on Day 2 in Fig.~\ref{fig:loadforecast}.
The accuracy of load forecasts is highly dependent on the accuracy of weather forecasts, which utility companies usually obtain from vendors.
The uncertainty in the exogenous factors must be accounted for in the final forecast, and because several of those factors are forecasts themselves, errors can be large. 
In this case, the model's performance is sufficient to add value to business operations at normal operating levels and in typical seasonal weather. The accuracy will be reduced during time of extreme cold or heat due to the lack of historical meter data.

A special case in load forecasting is cold load characterization. 
During steady state, the heating or cooling load on a feeder is typically a smaller percentage of the total heating or cooling load.  This reduced load results from the diversity of HVAC (heating, ventilation, and air conditioning) units simultaneously running due to normal cycling between on and off.  After an extended outage the temperature in the residence will likely fall outside the setpoint range.  Once the power is restored to the feeder, diversity of the heating or cooling load would be lost due to all the units turning on at the same time. This increase in load is referred to as "cold load". After some time period passes, the diversity will be restored because the unit run times will vary depending on factors such as HVAC rating, home size, and temperature setpoints. 
Cold load peak values are affected by pre-outage load behavior, season (winter/summer), time of day, ambient temperature, and load composition (customer types). 
Predicting these values at feeder breakers or other downstream protective devices enables optimal sequencing of operations to restore power quickly while minimizing the likelihood of damaging equipment. 
In addition, EMS typically has a load shed/restoration tool that can automatically conduct outage rotations among all feeders in the system during a short supply situation such as the recent Texas power crisis~\cite{wu2021open}.
With predictions of each feeder's post-outage load peaks, the EMS can automatically and accurately follow ISO's load-shed requirements to protect the entire power grid.  

Oncor is currently testing a linear regression model to predict the ratio of the peak cold-load (post-outage) and pre-outage load of a feeder. The data used are outage duration, pre- and post-outage temperatures, and the fraction of residential customers on the feeder. 
The residential load fraction is a good proxy for feeder load diversity ({i.e.}, the independently controlled cyclic loads such as HVAC systems that may be energized at any given time during normal operating conditions). 
Since feeder breaker level outages are relatively rare, feeders are grouped by their residential fractions and a model is learned for each feeder-group. 
A total of 1127 breakers were evaluated and training data were collected for fitting the regression model.
To accurately capture the cold load behavior, switch operation logs and fuse level events were reviewed to ensure that the cold load peaks were neither overestimated due to switching operations nor underestimated due to fuse level events behind the breakers.
During an emergency situation, this model will take the pre-selected outage durations for feeder rotations and post-outage temperatures as inputs. 
The model will output a predicted load ratio for each (phase) feeder and the power ratio, then the cold-load peaks can be estimated. These four predictions are useful for unbalanced feeders; in balanced feeders, a single estimate of the power ratio is sufficient.

\begin{figure}[tb]
\centering
\includegraphics[width=0.48\textwidth]{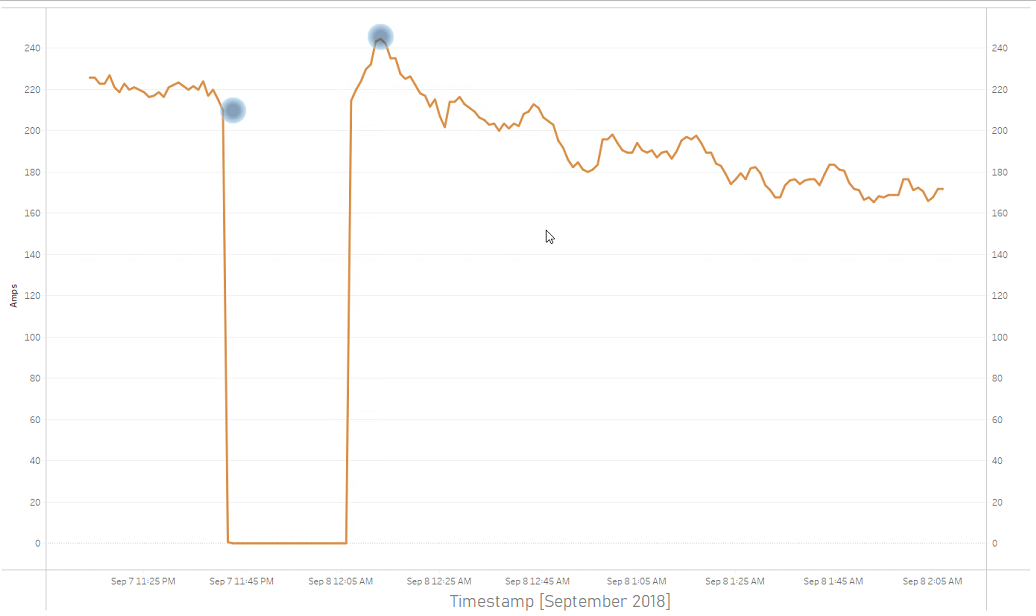}
\caption{An example of the cold load peek forecasting result for one feeder. Orange curve: SCADA current time series data before and after a feeder level outage; grey points: pre-outage current reading and predicted post-outage current.}
\label{fig:coldload}
\end{figure}

Fig.~\ref{fig:coldload} shows an example of the cold load peak prediction for one feeder using the trained regression model. The two highlighted points in the figure mark the pre-outage current and predicted post-outage current for one phase of a feeder. The predicted value is marked at the same location as the post-outage load peak only for better visualization and easier comparison. 

\subsection{Residential/Industrial/Commercial (RIC) Categorization}

For many transmission and distribution planning models, RIC percentages at each substation transformer bank are used to allocate load in the base-case models. 
These percentages are also used to derive the number of various motor types for dynamics models and simulations. 
Likewise, distribution planners must sometimes perform weather corrections for load projections.
In these cases, industrial and other non-weather-sensitive loads (such as water pumping and/or oil field pumping loads) are not weather-corrected, because these load types are rarely weather-sensitive or weather-dependent.
Traditionally, the customer category of a premise is established at the creation of the premise and may not get updated when the customer type changes. 
For example, a commercial building can be leased to a new business that has a completely different load profile from that of the previous business, but the utility may not be aware of the change.  

Before system-wide installation of advanced meters, the RIC process used typical summer and winter hourly load profiles for each category of the building distribution feeder models. 
With the availability of AMI interval data and distributed computing, the process can be improved by directly analyzing the load profile of each premise.
Because residential meters can usually be identified using information provided by ISOs, it is more valuable to focus on the non-residential meters.

As an initial approach, cluster analysis was conducted on data from approximately~$490,000$ non-residential meters. 
Domain experts selected 12~weeks (non-contiguous) over a 1-year period that adequately covered different seasonal and holiday effects (\emph{e.g.}, extended hours during holidays). 
The 15~minute interval load data was collected from each week and the time-series for each meter were stacked into 8064-dimensional vectors ($12\times 7\times 24\times 4$). 
$K$-means clustering was applied to the data with, initially, $k=100$. 
The initial parameter values were chosen as subject matter experts' estimation.
Subsequently, large clusters were checked by comparison of random samples within the cluster to the cluster center (i.e., comparing the average load profile with the other profiles within the cluster). 
If a large deviation was found, then the cluster was split. 
A less heuristic approach would be to use the $V$-measure or silhouette-coefficient to determine an optimal number of clusters~\cite{rosenberg2007v, ROUSSEEUW198753}; however, cluster-splitting was found to be effective for this use case.

This cluster analysis was conducted using Spark; 3--4 hours were needed for a cluster with 2 namenodes (dual Xeon-4208, 768 GB RAM per node) and 8 datanodes (dual Xeon-5218, 768 GB RAM per node) as shown in Table~\ref{tab:usecases}. The analysis will be repeated annually to capture any premises with changes in load type.

\subsection{Other Use Cases}
Many companies in the power industry have been developing data-driven methods for their business needs. Exelon Utility and ComEd applied classification methods to aerial/satellite images as well as light detection and ranging (LiDAR) data for vegetation management to better understand the system's tree trimming workload in the system seeking to cut rimming costs while reducing the number of tree-related outages~\cite{ComEd_LiDAR,ComEd_vegetation}. ISO New England proposed a prediction method based on decision tree to instruct interface limit values for different operating conditions~\cite{zhao2018enhanced}. Researchers in Hitachi proposed a three-layer wind power prediction model based on the data from historical power measurements and numerical weather prediction tools~\cite{gao2020three}. In addition, Bhattarai \emph{et al.} reviewed related literature on big data analytics from the perspectives of electric utilities and industry~\cite{Bhattarai2019Big}.

\section{Outlook}\label{sec:outlook}
In this paper, we have briefly reviewed the structure of power system physical and market operation, today's AI infrastructure of data acquisition and computation in power systems, state-of-the-art AI-based approaches for multiple critical functions, and industrial use cases of AI methods. In the following, we propose several research directions from the aspects of data, computing and AI algorithms.

\subsection{High-quality Open-source Datasets}
Despite the advances in data acquisition, in contrast to numerous datasets that have benefited broad AI communities, the lack of publicly accessible high-quality power datasets may be impeding the advancement of AI research in power systems. There are several reasons for the limited public access to power datasets. First, most real-world operational data are protected by policies such as CEII in the interest of confidentiality. Second, due to the reliability of real-world power grids, the rairty of opportunities to observe high-impact events may produce an insufficiently robust real-world measurement dataset. Third, the value of creating comprehensive and trustworthy benchmark power datasets has been overlooked by the power system community. There have been few open-source datasets~\cite{maslennikov2016test,zheng2021psml} and online contests dedicated to topics such as forced oscillation localization~\cite{OLS} and power system operation~\cite{marot2020learning,marot2021learning}. However, far more will be needed to build a standard library of open-source benchmark datasets along with critical tasks in clear mathematical formulation that can be used to train, calibrate, test and benchmark data-driven models. One critical challenge is that commonly used random sampling and data generation methods do not guarantee representativeness~\cite{joskow2019challenges} and may introduce unexpected biases into subsequent data-drive methods. Therefore, it is critical to investigate data generation methods that guarantee comprehensiveness and representativeness; these may be dataset-inspired or task-tailored. In the meantime, it is also necessary to propose algorithm-agnostic metrics to consistently assess the property of representativeness.

\subsection{Advanced Computing}
As mentioned in Section~\ref{sec:review of operation}, complex control algorithms are too time-consuming for real-time security control, especially in contingency scenarios. The rapid expansion of sensors has enabled massive data acquisition; however, although this data is necessary for realizing a digitized power grid, using all of it is beyond current computing capacity for centralized methods. Therefore, to explore and exploit advanced algorithms and massive streaming data, hybrid edge and cloud computing are necessary to dynamically balance the computational-load and escalate computing power as needed. For example, edge devices can compute partial results across several hundred sensors (e.g., half of a neural network's layers) and forward the results to the control center for final computations, effectively distributing computational load. Furthermore, new ASIC devices, dedicated to power system computations could be used in edge devices for real-time data processing and to accelerate simulations. In addition, communications between edge and cloud may contain sensitive information, requiring privacy preserving methods such as federated learning~\cite{mcmahan2017communication}.

Besides accelerating computation, platforms are needed to manage the complexity introduced by digitization. The software development industry uses a set of (automation) practices called ``DevOps'' to manage development, integration, testing, deployment, and monitoring of distributed software systems. In sectors where data-driven and machine learning algorithms are used, another layer is added to DevOps~\cite{ebert2016devops,zhu2016devops} that encompasses automated training, testing, deployment, and monitoring of models---this is called ``MLOps"~\cite{NIPS2015_86df7dcf, treveil2020introducing}. Both DevOps and MLOps lower the maintenance cost of complex software systems through automation, but the initial investment is high. For efficient digitization of the power grid, both DevOps and MLOps will be necessary; however, there are unique aspects of power systems that require investigation. Because the grid is primarily hardware, it would be highly imprudent to blindly adopt methods developed for pure software environments.

The instrumentation and sensors being deployed into modern grids also bring cyber-security challenges. If the data and contols are transmitted over the internet (\emph{e.g.}, cloud computing), the grid is vulnerable to the same cyber-attacks as a website, except the stakes are much higher: outages, energy theft, and loss of private data. Monitoring and detecting cyber-threats to the grid is an important area for cross disciplinary research combining power systems, cyber-security, and AI. 

\subsection{Use-inspired AI Methods for Practical Applications}

Because power grids are large-scale critical infrastructure systems for human society, future research efforts ought to use-inspired AI algorithms that possess three key properties, namely interpretability, robustness, and scalability, aiming to facilitate practical applications. First, AI algorithms ought to be explainable by first-principle-based physical models, because only interpretable algorithms are acceptable for participation in the human-in-the-loop decision making process. In particular, interpretable AI approaches should provide clear causal inference for the purposes of real-time monitoring, control and diagnosis, such as identifying root cause of complex observations. Preliminary efforts have been devoted to physics-informed ML as summarized in~\cite{karniadakis2021physics}. The principle is to steer the learning process towards identifying physically consistent solutions, of which instructive guidance contains three aspects, namely data processing, loss function modification, and model architecture design. For example, incorporating ordinary different equation (ODE) formats into loss function as regularization terms can improve the performance of system identification algorithms based on transient data or improve the fidelity of transient data generation methods. Second, AI algorithms must have performance guarantees extending beyond the basic, unperturbed scenarios. Particularly, the robustness to perturbation is critically important for reinforcement learning-based algorithms for decision making. Meta reinforcement learning~\cite{nagabandi2018learning,arndt2020meta} and transfer learning can potentially accommodate the gap between reality and simulation environment, thereby rendering the decision making adaptive to varying conditions and scenarios. Third, another highly desirable feature of AI algorithms is scalability, which refers to adequate effectiveness and efficiency in large-scale real-world systems. The concern regarding scalability arises from the aforementioned observation that the performance of existing AI algorithms in the power system domain is mostly demonstrated by small-scale grids without validation in large-scale cases. As high-dimensional measurements in power systems empirically have properties such as approximate low-rankness and sparsity, they may be potentially efficacious to discover intrinsic low-dimensional manifolds and linear coordinates in data structure~\cite{brunton2016discovering}.

In summary, digitization of the power grid will play a major role in transforming the electricity sector into a decarbonized system while simultaneously improving grid reliability.  The synergy of high-dimensional dynamic data, increased computing power, and use-inspired AI algorithms, will enable improvements to the reliability and operational efficiency of the power grid at multiple scales. Challenges remain on the integration of heterogenous data sets, cyber-physical security, and development of robust, interpretable AI algorithms. 
Strong collaboration between industry and academia will be crucial for the successful adoption of use-inspired AI methods in a decarbonized power system.


\section*{Acknowledgements}
The authors sincerely thank Jimmy Liu, Steven Dennis, and Thomas Wilson for their help on the Oncor use cases presented in this paper.

\bibliographystyle{IEEEtran}
\bibliography{main}

\section*{Authors}
\begin{IEEEbiography}
[{\includegraphics[width=1in,height=1.25in,clip,keepaspectratio]{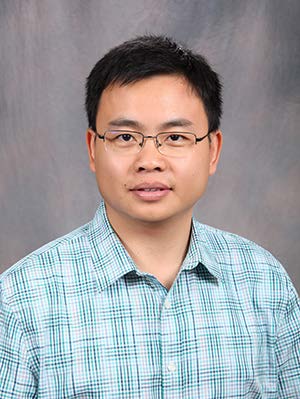}}] 
{Le Xie} (Fellow, IEEE) received his B.E. degree in Electrical Engineering from Tsinghua University, Beijing, China, in 2004, an M.S. degree in Engineering Sciences from Harvard University,
Cambridge, MA, USA, in 2005, and a Ph.D. degree from Carnegie Mellon University, Pittsburgh, PA,
USA, in 2009. He is currently a professor with the Department of Electrical and Computer Engineering, Texas A\&M University, College Station, TX, USA. His research interests include modeling and control of large-scale complex systems, smart grids application with renewable energy resources, and electricity markets.
\end{IEEEbiography}

\begin{IEEEbiography}
[{\includegraphics[width=1in,height=1.25in,clip,keepaspectratio]{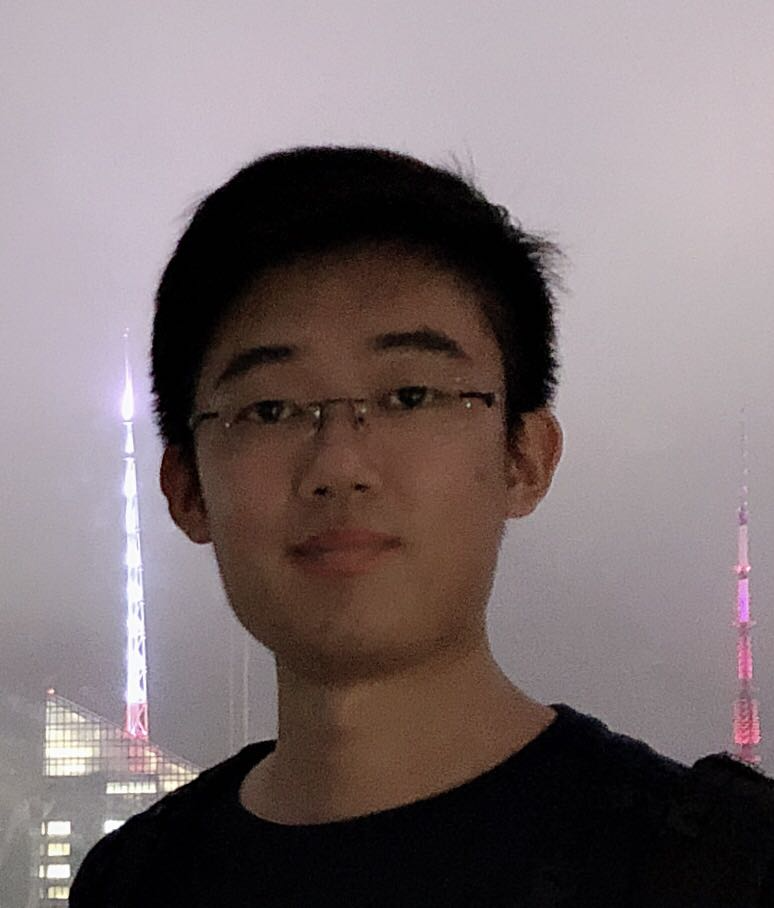}}] 
{Xiangtian Zheng} (Graduate Student Member, IEEE) received his B.E. degree from Tsinghua University, Beijing, China, in 2017. He is currently pursuing a Ph.D. degree in Electrical Engineering at Texas A\&M Univerisity. His industry experience includes an internship with PJM in 2019 and an internship with Mitsubishi Electric Research Laboratory in 2021. His research interests include domain knowledge-informed machine learning for power system security.
\end{IEEEbiography}

\begin{IEEEbiography}
[{\includegraphics[width=1in,height=1.25in,clip,keepaspectratio]{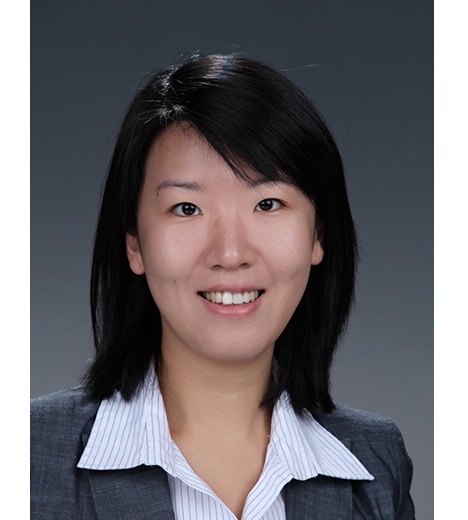}}] 
{Yannan Sun} (Senior member, IEEE) received her B.S. degree in Mathematics from the University of Science and Technology of China, Hefei, China, in 2004, and an M.S. degree in Statistics and a Ph.D. degree in Mathematics from Washington State University, Pullman, WA, USA, in 2007 and 2010, respectively. She was a Scientist/Senior Scientist in the Electricity Infrastructure group at Pacific Northwest National Laboratory, Richland, WA, USA, from 2010 to 2017. She is currently a data scientist at Oncor Electric Delivery. Her expertise lies in data analytics and machine learning using power system data, which she has employed to develop many data-driven algorithms for load forecasting, anomaly detection, connectivity correction and equipment preventive maintenance.
\end{IEEEbiography}

\begin{IEEEbiography}
[{\includegraphics[width=1in,height=1.25in,clip,keepaspectratio]{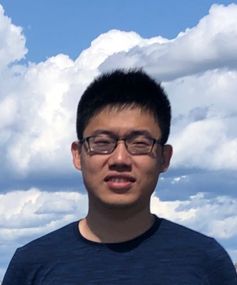}}] 
{Tong Huang} (Member, IEEE) is a postdoctoral researcher at Massachusetts Institute of Technology (MIT). Before joining MIT, he was a postdoctoral researcher at  Texas A\&M University in 2021. He received his B.E. degree in Electrical Engineering from North
China Electric Power University in 2013 and an M.S. and a Ph.D. degree in Electrical Engineering from Texas A\&M University in 2017 and 2021, respectively. He was a Visiting Student with the Laboratory for Information and Decision Systems, MIT, in 2018. His industry experience includes an internship with ISO-New England in 2018 and an internship with Mitsubishi Electric Research Laboratories in 2019. He received the Best Paper Award at the 2020 IEEE PES General Meeting, the Best Paper Award at the 54-th Hawaii International Conference on System Sciences, Thomas W. Powell'62 and Powell Industries Inc. Fellowship, and Texas A\&M Graduate Teaching Fellowship.
\end{IEEEbiography}

\begin{IEEEbiography}
[{\includegraphics[width=1in,height=1.25in,clip,keepaspectratio]{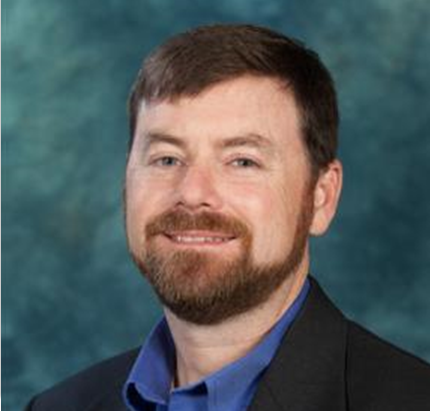}}] 
{Tony Bruton}
(Member, IEEE) received his B.S. degree in Electrical Engineering from Texas Tech University in 2000.  He began working for Oncor as a substation design engineer. For several years, he managed Oncor’s high voltage grid in east Texas then managed the group that designed and built high voltage transmission lines.  He also managed the routing and acquisition of right-of-way for new Transmission lines in west Texas.  Mr. Bruton’s current role as Director of T\&D Services involves ensuring an accurate system that monitors and controls Oncor’s Transmission and Distribution systems, EMS and ADMS, as well as data analytics and operator training.
\end{IEEEbiography}

\end{document}